%% file: main.tex
\documentclass[apj,twocolumn,numberedappendix]{openjournal}
\input{Packages/packages.tex}

\begin{document}

\newcommand{\cmt}[1]{\textcolor{red}{\textsf{#1}}}
\newcommand{\CE}[1]{\textcolor{cyan}{$\langle$\textsf{CE: #1}$\rangle$}}
\newcommand{\GC}[1]{\textcolor{green}{$\langle$\textsf{GC: #1}$\rangle$}}
\newcommand{\MKP}[1]{\textcolor{magenta}{$\langle$\textsf{MKP: #1}$\rangle$}}
\newcommand{\SM}[1]{\textcolor{orange}{$\langle$\textsf{SM: #1}$\rangle$}}
\newcommand{\REV}[1]{\textcolor{red}{$\langle$\textsf{Rev: #1}$\rangle$}}

\renewcommand{\d}{\mathrm{d}}

\defcitealias{EverettCotter_2025}{Paper I}
\defcitealias{PaperIII}{Paper III}

\journalinfo{The Open Journal of Astrophysics}

\title[]{DIPLODOCUS II: Implementation of transport equations and test cases relevant to micro-scale physics of jetted astrophysical sources}
\shorttitle{DIPLODOCUS II}

\author{\vspace{-1.5cm}Christopher N. Everett$^{1,\star}$\orcidlink{0000-0001-5181-108X}, Marc Klinger-Plaisier$^{2}$\orcidlink{0000-0002-4697-1465}, Garret Cotter$^{1}$\orcidlink{0000-0002-9975-1829}}
\affiliation{$^1$Oxford Astrophysics, Denys Wilkinson Building, Keble Road, Oxford, OX1 3RH, United Kingdom}
\affiliation{$^2$Anton Pannekoek Institute for Astronomy, University of Amsterdam, Science Park 904, 1098 XH Amsterdam, The Netherlands}
\thanks{Contact e-mail:  \href{mailto:christopher.everett@physics.ox.ac.uk}{christopher.everett@physics.ox.ac.uk}}



\begin{abstract}
    \input{Sections/abstract.tex}
\end{abstract}


\maketitle

\input{Sections/introduction.tex}
\input{Sections/prelimiary.tex}

\input{Sections/collisions.tex}
\input{Sections/transport.tex}
\input{Sections/test_cases.tex}

\input{Sections/conclusion.tex}

\section*{Acknowledgements}
C.\,N.\,Everett acknowledges a Science and Technologies Facilities Council (STFC) studentship ST/X508664/1. G.\,Cotter acknowledges support from STFC grants ST/V006355/1, ST/V001477/1 and ST/S002952/1, and from Exeter College, Oxford. M.\,Klinger-Plaisier is supported by the European Research Council (ERC) Synergy Grant ``BlackHolistic:  Colour Movies of Black Holes: Understanding Black Hole Astrophysics from the Event Horizon to Galactic Scales’' (grant 10107164). The authors would also like to thank James Matthews, Sera Markoff (and their research group), and Dion Everett for their helpful insights during various phases of this work.

\section*{Data Availability}

The simulation data presented in this paper are available from the authors upon reasonable request. Data was generated using \texttt{Diplodocus.jl} v.0.1.1 and \texttt{AM$^3$} v.1.0.1, both of which are open source. Documentation can be found at their respective websites \href{https://cneverett.github.io/Diplodocus.jl/}{\texttt{Diplodocus.jl}} and \href{https://am3.readthedocs.io/en/latest/}{\texttt{AM$^3$}}.

\bibliographystyle{mnras_modified}
\bibliography{references}

\onecolumngrid
\appendix

    \input{Appendicies/differential_cross_sections.tex}

    \input{Appendicies/radreactandsync.tex}
    \input{Appendicies/scattering_limits.tex}
    \input{Appendicies/minkowski_coordinates.tex}
    \input{Appendicies/test_cases_num_eng.tex}

    \input{Appendicies/obs_flux.tex}

\end{document}

%% file: Packages/packages.tex
\usepackage[utf8]{inputenc}
\usepackage[T1]{fontenc}
\usepackage{ae,aecompl}

\usepackage{graphicx}
\graphicspath{Figures/}
\usepackage{amsmath,amsthm,amssymb,amsfonts,amstext}
\usepackage{bm}
\usepackage{orcidlink}

\usepackage{enumitem}

\usepackage{booktabs}
\usepackage[ruled]{algorithm2e}

\usepackage{savesym}
\savesymbol{tablenum}
\usepackage{siunitx}
\restoresymbol{SIX}{tablenum}

\usepackage{float}

\newcommand{\aref}[1]{Appendix~\ref{#1}}
\newcommand{\asref}[1]{Appendices~\ref{#1}}
\newcommand{\eref}[1]{Eq.~(\ref{#1})}
\newcommand{\esref}[1]{Eqs.~(\ref{#1})}
\newcommand{\fref}[1]{Fig.~\ref{#1}}
\newcommand{\fsref}[1]{Figs.~\ref{#1}}
\newcommand{\sref}[1]{Section~\ref{#1}}
\newcommand{\ssref}[1]{Sections~\ref{#1}}

\usepackage{hyperref}
\hypersetup{colorlinks=true,linkcolor=blue,citecolor=blue,filecolor=blue,urlcolor=blue}
\usepackage[all]{hypcap}

%% file: Sections/abstract.tex
DIPLODOCUS (\textbf{D}istribution-\textbf{I}n-\textbf{PL}ateaux meth\textbf{ODO}logy for the \textbf{C}omp\textbf{U}tation of transport equation\textbf{S}) is a framework being developed for the mesoscopic modelling of astrophysical systems via the transport of particle distribution functions through the seven dimensions of phase space, including continuous forces and discrete interactions between particles. Following \citetalias{EverettCotter_2025}, which details the mathematical background, this second paper provides an overview of the numerical implementation in the form of the code package \href{https://cneverett.github.io/Diplodocus.jl/}{\texttt{Diplodocus.jl}}, written in \texttt{Julia}, including the description of a novel Monte-Carlo sampling technique for the pre-computation of anisotropic collision integrals. In addition to the discussion of numerical implementation, a selection of test cases are presented to examine the package's capabilities. These test cases focus on micro-scale physical effects: binary collisions, emissive interactions, and external forces that are relevant to the modelling of jetted astrophysical sources, such as Active Galactic Nuclei and X-Ray Binaries.

%% file: Sections/introduction.tex
\section{Introduction}\label{sec:intro}

DIPLODOCUS is a framework for particle transport through phase space, being developed to assist in improving our understanding of a wide range of astrophysical sources by relating that particle transport to observable emission spectra. 

The context and driving reasoning for this work is detailed in the introduction of the first paper in this series, hereby referred to as \citetalias{EverettCotter_2025}. In addition to the context of this work, \citetalias{EverettCotter_2025} describes the mathematical framework in detail. Therefore the reader is encouraged to consult that paper for definitions and clarification of terms not reproduced here. This second paper in the series focuses on the numerical implementation and testing of the framework in the form of the open source package \texttt{Diplodocus.jl}\footnote{Documentation, with tutorials, can be found at: \href{https://cneverett.github.io/Diplodocus.jl/}{Diplodocus.jl}} .

\texttt{Diplodocus.jl} is written in \texttt{Julia} \citep{Julia-2017} and is split into three sub-packages. \texttt{DiplodocusCollisions.jl} performs the evaluation of collision matrices via Monte-Carlo integration (as described in \sref{sec:collisions}). These collision matrices define the rate at which particles are created/destroyed/transferred between different regions of momentum space. These collision matrices may then be loaded into \texttt{DiplodocusTransport.jl} which simulates the time evolution of those particles over the whole of phase space, including the effects of continuous forces which may be conservative or non-conservative (as described in \sref{sec:transport}). Results of these simulations can then be viewed using a suite of plotting functions hosted by the \texttt{DiplodocusPlots.jl} sub-package, which builds upon the \texttt{Makie.jl} \citep{DanischKrumbiegel2021} plotting ecosystem, allowing for the generation of static, animated and interactive plots.

Following the descriptions of the numerical implementation in \ssref{sec:collisions} and \ref{sec:transport}, test cases of micro-scale physics, that is interactions and forces that affect how particles are distributed in momentum-space, are presented in \sref{sec: test cases}\footnote{Tests of macro-scale effects i.e. those which affect how particles are distributed in physical space, are not included in this work, instead being the focus of \citetalias{PaperIII}, which will also include discussions regarding the \texttt{Diplodocus.jl}'s performance and scaling to high power computing systems.}. 

At the micro-physics level, \texttt{Diplodocus.jl} has been designed to handle a wide range of effects, including: binary collisions, external forces, both conservative and non-conservative and emissive interactions. The test cases presented in \sref{sec: test cases} have been designed to be relevant to the modelling of jetted astrophysical sources, while still demonstrating the code's capabilities. \sref{subsec: hard spheres} examines the thermalisation and isotropisation of a population of elastically colliding spheres. \sref{subsec: ExB} examines the gyration and drift of electrons under the influence of uniform magnetic and electric fields. \sref{subsec: ele rad raction} considers the non-conservative force of radiation reaction, observing the cooling of electrons and the phenomenon of population inversion. \textit{Radiation} reaction implies the presence of a radiated photon population; \sref{subsec: AM3 sync} examines synchrotron radiation emitted by cooling electrons and provides a direct comparison between \texttt{Diplodocus.jl} and the single-zone code \texttt{AM\textsuperscript{3}}. Further, \sref{subsec: AM3 SSC} adds in (inverse-)Compton scattering to assess the process of Synchrotron Self-Compton (SSC) and finally, \sref{subsubsec: ani ssc} introduces anisotropy to the SSC model by the inclusion of a helical magnetic field structure, demonstrating variation in the observed spectra that is likely to affect parameter fitting of jetted sources including blazars.

%% file: Sections/prelimiary.tex
\section{Numerical Preliminaries}\label{sec: prelim}
\subsection{Distribution-In-Plateaux}\label{subsec: DIP}
Under the \textit{Distribution-In-Plateaux} (DIP) formalism described in \citetalias{EverettCotter_2025}, the particle distribution function in phase space $f(\bm{x},\bm{p})$ takes the form:
\begin{equation}
\begin{split} \label{eqn: DIP}
    f(\bm{x},\bm{p}) =& \sum_{\alpha\beta\gamma\delta ijk} \frac{f_{\alpha\beta\gamma\delta ijk}}{p^2\Delta p_i\Delta u_j\Delta\phi_k} H_{\alpha}(t)H_{\beta}(x)
    \\
    &\times H_{\gamma}(y)H_{\delta}(z)H_{i}(p)H_{j}(u)H_{k}(\phi),
\end{split}
\end{equation}
where $\{x,y,z\}$ refer abstractly to a set of three spatial coordinates with Greek subscripts referring to discrete domains of those coordinates, $\{p,u,\phi\}$ are the modified spherical coordinates ($u=\cos\theta$) used as momentum space coordinates\footnote{The orientation of the spherical momentum-space coordinates with respect to the spatial coordinates is determined by coordinate transform $e_a^{~\alpha}$. See \aref{app: coords} and \citetalias{EverettCotter_2025} for more details.}, with Latin subscripts referring to discrete domains of those coordinates. The function $H$ is a boxcar function where, for example, $H_\alpha(t)$ is defined as
\begin{align}
\begin{split} \label{eqn: DIP Boxcar}
    H_{\alpha}(t)&=\Theta(t-t_\alpha) - \Theta(t-t_{\alpha+1})
    \\
    &=
    \begin{cases}
        1, & \text{for } t_{\alpha}<t<t_{\alpha+1},
        \\
        h_\alpha^- & \text{for } t=t_{\alpha} 
        \\
        h_\alpha^+=1-h_\alpha^- & \text{for } t=t_{\alpha+1} 
        \\
        0, & \text{for all other } t,
    \end{cases}
\end{split}
\end{align}
where $h_\alpha^\pm$ are the values of the boxcar function at the bounds of the coordinate sub-domain $t_\alpha\le t \le t_{\alpha+1}$, left as a variable to allow for modifications to the numerical scheme (see \sref{subsec: scheme}).

\subsection{Units and Normalisation}\label{subsec: units norm} 
Internally, the code normalises momentum by the product of the electron rest mass and the speed of light: $m_ec \times p\,[\text{code units}]=p\,[\unit{\kg\meter\second^{-1}}]$, and the time coordinate $t$ by the product of the Thomson scattering cross section and speed of light: $t\,[\text{code units}]=  \sigma_\text{T} c \times t\,[\text{s}]$. The latter corresponds to the characteristic interaction time scale between particle distributions with number density of $1\,\text{m}^{-3}$ and interaction cross section $\sigma_\text{T}$. Equations presented in this paper are given in un-normalised SI units.

%% file: Sections/collisions.tex
\section{Collisions}\label{sec:collisions}
Within the DIPLODOCUS framework, the effects of collisions are included via collision matrices. These matrices describe the transfer of particles between different momentum-space sub-domains as a result of discrete interactions between particles. The evaluation of these matrices is performed by functions contained within the \texttt{DiplodocusCollisions.jl} sub-package of \texttt{Diplodocus.jl}.

\subsection{Binary Collisions}\label{subsec: binary collisions}
The gain and loss array elements for reversible binary interactions $\mathfrak{ab}\leftrightharpoons\mathfrak{cd}$ are in general given by Eqs.~(A5) to (A8) of \citetalias{EverettCotter_2025}. These equations, with the same terminology as \citetalias{EverettCotter_2025}, are  reproduced for convenience in \esref{eqn: binary gain matrix} to \eqref{eqn: binary loss and gain terms}. Within these equations $G_{\mathfrak{a}\mathfrak{b}\rightarrow\mathfrak{c}\mathfrak{d},ijklmnopq}$ is an element of a nine dimensional array describing the \textit{gain} of particles of type-$\mathfrak{c}$ into the momentum space sub-domain $P_{ijk}$ from a binary interaction between particles of type-$\mathfrak{a}$ in the sub-domain $P_{lmn}$ and particles of type-$\mathfrak{b}$ in sub-domain $P_{opq}$ and $L_{\mathfrak{a}\mathfrak{b}\leftarrow\mathfrak{c}\mathfrak{d},ijklmn}$ is an element of a six dimensional array describing the \textit{loss} of particles of type-$\mathfrak{c}$ in sub-domain $P_{ijk}$ due to interactions with particles of type-$\mathfrak{d}$ in sub-domain $P_{lmn}$. 

\begin{widetext}
\begin{subequations}
    \begin{align}
    \label{eqn: binary gain matrix}
    G_{\mathfrak{a}\mathfrak{b}\rightarrow\mathfrak{c}\mathfrak{d},ijklmnopq}&=\int_{P_{ijk}}\int_{P_{lmn}}\int_{P_{opq}}G_{\mathfrak{a}\mathfrak{b}\rightarrow\mathfrak{c}\mathfrak{d}}\frac{\mathrm{d}^3p_\mathfrak{b}}{(p_\mathfrak{b})^2\Delta p_{\mathfrak{b},o}\Delta u_{\mathfrak{b},p}\Delta \phi_{\mathfrak{b},q}}\frac{\mathrm{d}^3p_\mathfrak{a}}{(p_\mathfrak{a})^2\Delta p_{\mathfrak{a},l}\Delta u_{\mathfrak{a},m}\Delta \phi_{\mathfrak{a},n}}\mathrm{d}^3p_\mathfrak{c}, 
    \\
    \label{eqn: binary loss matrix}
    L_{\mathfrak{a}\mathfrak{b}\leftarrow\mathfrak{c}\mathfrak{d},ijklmn} &= \int_{P_{ijk}}\int_{P_{lmn}}L_{\mathfrak{a}\mathfrak{b}\leftarrow\mathfrak{c}\mathfrak{d}}\frac{\mathrm{d}^3p_\mathfrak{d}}{(p_\mathfrak{d})^2\Delta p_{\mathfrak{d},l}\Delta u_{\mathfrak{d},m}\Delta \phi_{\mathfrak{d},n}}\frac{\mathrm{d}^3p_\mathfrak{c}}{(p_\mathfrak{c})^2\Delta p_{\mathfrak{c},i}\Delta u_{\mathfrak{c},j}\Delta \phi_{\mathfrak{c},k}},
    \\
    \label{eqn: binary loss and gain terms}
    G_{\mathfrak{a}\mathfrak{b}\rightarrow\mathfrak{c}\mathfrak{d}}&=\frac{\delta\left(p^0_\mathfrak{a}+p^0_\mathfrak{b}-p^0_\mathfrak{c}-p^0_\mathfrak{d}\right)}{\left(1+\delta_{\mathfrak{ab}}\right)\left(1+\delta_{\mathfrak{cd}}\right)}\frac{\mathcal{F}_{\mathfrak{a}\mathfrak{b}}^2}{\pi p^0_\mathfrak{a}p^0_\mathfrak{b}p^0_\mathfrak{c}p^0_\mathfrak{d}}\frac{\mathrm{d}\sigma_{\mathfrak{a}\mathfrak{b}\rightarrow\mathfrak{c}\mathfrak{d}}}{\mathrm{d}T}, 
    \quad 
    L_{\mathfrak{a}\mathfrak{b}\leftarrow\mathfrak{c}\mathfrak{d}} = \frac{1}{\left(1+\delta_{\mathfrak{cd}}\right)}\frac{\mathcal{F}_{\mathfrak{c}\mathfrak{d}}\sigma_{\mathfrak{a}\mathfrak{b}\leftarrow\mathfrak{c}\mathfrak{d}}}{p^0_\mathfrak{c}p^0_\mathfrak{d}},
    \end{align}
\end{subequations}
\end{widetext}

Evaluation of these array elements is performed using Monte-Carlo integration. This works by randomly sampling points in the integration domain, evaluating the value of the integration function at those points and averaging by the total number of points sampled to generate an approximation of this integral. To illustrate this concept in a lower number of dimensions, consider a generic 2D function $F(x,y)$\,---\,the Monte-Carlo integral estimate of this function over the sub-domains $\Delta x$ and $\Delta y$ takes the form:
\begin{equation}\label{eqn: MC int}
    \int_{\Delta x}\int_{\Delta y} F(x,y)\mathrm{d}x\mathrm{d}y\approx \frac{\Delta x \Delta y}{N}\sum_{\alpha=1}^{N}F(x_\alpha,y_\alpha).
\end{equation}
The error estimate for Monte-Carlo integration scales as $\mathcal{O}(N^{-1/2})$, independent of the number of dimensions $d$ of the integral, whereas grid-based quadrature methods scale as $\mathcal{O}(N^{-k/d})$, where $k$ is the order of the scheme being used \citep{hammersleyMonteCarloMethods1964,caflischMonteCarloQuasiMonte1998}. Therefore, when the dimensionality is large, e.g. \esref{eqn: binary gain matrix} and \eqref{eqn: binary loss matrix}, the Monte-Carlo methods become the only practical way to evaluate integrals\footnote{The use of low-discrepancy sequences to generate quasi-random points can be used to further improve the convergence rate of Monte-Carlo integration \citep{caflischMonteCarloQuasiMonte1998}. However, the use of such sequences is not currently implemented within \texttt{DiplodocusCollisions.jl}.}.

For the loss array (\esref{eqn: binary loss matrix} and \eqref{eqn: binary loss and gain terms}), the process of Monte-Carlo integration is implemented as follows: first, momentum states of the incoming particles are sampled uniformly over momentum space $\{\bm{p}_\mathfrak{c},\bm{p}_\mathfrak{d}\}_\alpha$. Then their sub-domain location in momentum space is determined, and the value of the integration function $ L_{\mathfrak{a}\mathfrak{b}\leftarrow\mathfrak{c}\mathfrak{d}}$ is added to that sub-domain's total, and the tally of points sampled in that sub-domain $N$ incremented by one. The integration estimate is then given by:
\begin{equation}\label{eqn: binary loss matrix MC}
    L_{\mathfrak{a}\mathfrak{b}\leftarrow\mathfrak{c}\mathfrak{d},ijklmn}= \frac{1}{N_\text{loss}}\sum_{\alpha=1}^{N_\text{loss}}L_{\mathfrak{a}\mathfrak{b}\leftarrow\mathfrak{c}\mathfrak{d}}(\{\bm{p}_\mathfrak{c},\bm{p}_\mathfrak{d}\}_\alpha),
\end{equation}
which is implemented to be agnostic to the specific binary interaction, hence providing a single method for all binary interactions, with only the total cross section $\sigma_{\mathfrak{a}\mathfrak{b}\leftarrow\mathfrak{c}\mathfrak{d}}$ needed to be supplied.

For the gain array (\eref{eqn: binary gain matrix}), the process of integration is complicated by the Dirac delta function in the  $G_{\mathfrak{a}\mathfrak{b}\rightarrow\mathfrak{c}\mathfrak{d}}$ term. This results in one of the momentum variables being dependent on the others, prohibiting sampling of that variable directly. If care is not taken this can lead to poor accuracy of integration in the sub-domains of the dependent variable.

\subsubsection{Importance Sampling}\label{subsec: importance sampling}
\begin{figure}[!b]
    \centering
    \includegraphics[]{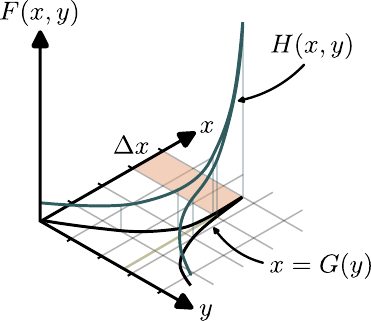}
    \caption{\label{fig: 2D MC analogy 1}A generic function $F(x,y)$ (\eref{eqn: MC test func}), with magnitude given by the function $H(x,y)$ and $x$ is a dependent variable constrained by $x=G(y)$. Accurate Monte-Carlo integration of this function over sub-domains $x$ (e.g. the sub-domain $\Delta x$) is difficult as $x$ cannot be directly sampled and a large contribution to the integral estimate can arise from a small domain $y$ values.}
\end{figure}

To describe the problem of how a dependent variable can lead to poor integration accuracy in more detail, consider again a generic 2D function $F(x,y)$, however, now the variable $x$ is dependent on $y$, being constrained by a Dirac delta function: 
\begin{equation}\label{eqn: MC test func}
    F(x,y)=\delta(x-G(y))H(x,y).
\end{equation}
If this function were to be plotted over the $xy-$plane it would appear as a line, rather than a surface, with the magnitude of the function being given by another generic function $H(x,y)$. The problem arises when a major contribution to the integration of $F(x,y)$ in some sub-domain of the dependent variable $x$ comes from a small region of the independent variable $y$ (\fref{fig: 2D MC analogy 1}). Therefore, if $y$ is sampled uniformly, this contributing region may be poorly sampled, leading to a poor integration estimate in that sub-domain of $x$.

In the context of binary collisions, this issue of poor sampling typically arises when one particle has significantly more momentum than the other, as the direction of an outgoing particle is not likely to be uniformly distributed on a unit sphere. Therefore, if the outgoing particle's momentum is taken to be the dependent variable, when sampling its direction of propagation it is unlikely that an important or even physical output state is sampled, leading to a poor integration estimate. 

When integrating a function that is known to be peaked in a small region using Monte-Carlo methods, a standard approach is to weight the sampling of points such that the peak of the function is well sampled, known as \textit{importance sampling}. Consider \eref{eqn: MC int} applied to \eref{eqn: MC test func}, but instead of sampling the independent variable $y$ uniformly it is sampled using a weighting function $y=w(Y)$, where $Y$ is a new, uniformly sampled, random variable, and $w(Y)$ is a monotonically increasing, differentiable function bounded by the limits of $y$ over the domain of $Y$. The integration estimate is then given by 
\begin{equation}\label{eqn: MC int is}
    \int_{\Delta x}\int_{\Delta y} F(x,y)\mathrm{d}x\mathrm{d}y\approx \frac{\Delta y }{N}\sum_{\alpha=1}^{N}\frac{H(G(y_\alpha),y_\alpha)}{P(y_\alpha)},
\end{equation}
where $G(y_\alpha)\in \Delta x$, $y_\alpha\in\Delta y$, $P(y)$ is the probability of sampling a point $y$ given by $P(y)=\int\delta(y-w(Y))\d Y$ and $N$ is the number of points sampled\footnote{As $x$ is a dependent variable, the value of $N$ is the total number of points sampled for which $y_\alpha\in\Delta y$, irrespective of if $G(y_\alpha)\in\Delta x$.}. 

In the context of Particle-In-Cell simulations, due to the finite number of particles, collisions integrals may only need be sampled close to their peaks, i.e. the most likely outgoing interaction states. This is typically achieved by numerically inverting the differential cross section to form a cumulative distribution function which acts as the weighting function with which to sample outgoing states \citep{levinson_particle--cell_2018,gaudioComptonScatteringParticleincell2020a,crinquand_multidimensional_2020,mehlhaff_kinetic_2023}. However, this has three main drawbacks. Primarily this process is computationally intensive due to iterative sampling of the cumulative distribution function. Secondly, as a cost of this computational intensity, only limited samples can be generated, biasing the outcome towards the most likely output states. As a consequence, any tail in the interaction spectrum may be missed, leading to poor sampling of sub-domains $\Delta x$ that lie outside the peak. Third, the functions used in sampling must be created and tailored for each specific interaction, making the introduction of a new interaction to any simulation a laborious task. 

As particle interactions are pre-computable within the DIPLODOCUS framework, the intensity of the techniques used in PIC is not a major issue. However there is no compelling reason to only obtain a good integration estimate only for the peak of a given interaction; instead collisions may be well sampled over the entire phase space of the outgoing state without additional simulation cost. Further, the required tailoring and generation of cumulative distribution function for specific interactions go against the spirit of generality and expandability the authors are attempting to achieve with the DIPLODOCUS. Due to these reasons, inversion of the differential cross section was not chosen as an integration strategy for the evaluation of collision integrals within \texttt{DiplodocusCollisions.jl}.

Rather than using a different weighting scheme for each binary collision, an adaptable weighting scheme has been developed that is agnostic to specific interaction. This probability of sampling the independent variable $y$, $P_{j}(y_\alpha,\rho_j)$, is now dependent on a parameter $\rho$, which may be scaled to adequately sample different integration domains of the dependent parameter $x$.

For a given $\rho_j$ the integration estimate given by \eref{eqn: MC int is} applied to \eref{eqn: MC test func} is:
\begin{equation}
\begin{split}
    I_j=\frac{\Delta y}{N_j}\sum_{\alpha=1}^{N_j}\frac{H(G(y_\alpha),y_\alpha)}{P_j(y_\alpha,\rho_j)},
\end{split}
\end{equation}
and let $k_j$ denote the number of those samples for which $G(y_\alpha)\in\Delta x$. If multiple integration estimates are generated (each using a different value of $\rho$), the total integral estimate may then be given by a weighted average:
\begin{equation}\label{eqn: weighted average}
    I=\frac{I_1k_1+I_2k_2+I_3k_3+...}{k_1+k_2+k_3+...}.
\end{equation}
where $k_j$ is used to the weight each integral estimate\footnote{The process of using a weighted average of different samplings is similar to a technique used to optimise Monte-Carlo rendering of computer generated images \citep{veachOptimallyCombiningSampling1995}.}. 

Storing $I_j$, $N_j$ and $k_j$ for many sampling schemes is impractical, therefore numerical integration is performed by re-writing \eref{eqn: weighted average} in a recursive form such that only the current state and next state are required at any given time:
\begin{equation}\label{eqn: weighting scheme}
    I=\frac{\left(I_1k_1+I_2k_2\right)+I_3k_3}{\left(k_1+k_2\right)+k_3}=\frac{\tilde{I}\tilde{W}+I_3k_3}{\tilde{W}+k_3},
\end{equation} where $\tilde{I}=\left(I_1k_1+I_2k_2\right)/\left(k_1+k_2\right)$, is the old integral estimate with a weight $\tilde{W}=k_1+k_2$. It can also be shown that this preserves the same final integral estimate independent of the order in which the weighted averaging is performed.

\subsubsection{Weighting Scheme}\label{subsec:weighting}
Taking the momentum of the outgoing particle $p_\mathfrak{c}$ as the dependent variable, the gain array (\esref{eqn: binary gain matrix} and \eqref{eqn: binary loss and gain terms}) may be integrated once analytically \citep{EverettCotter_2024} to produce \esref{eqn: binary gain matrix delta removed} and \eqref{eqn: binary gain term delta removed}, in which   
$p_\mathfrak{c}=p_\pm$ are the two roots of the Dirac delta function $\delta\left(p^0_\mathfrak{a}+p^0_\mathfrak{b}-p^0_\mathfrak{c}-p^0_\mathfrak{d}\right)$, which are a function of the independent variables $p_\mathfrak{a},u_\mathfrak{a},\phi_\mathfrak{a},p_\mathfrak{b},u_\mathfrak{b},\phi_\mathfrak{b},u_\mathfrak{c},\phi_\mathfrak{c}$, and $\cos\Theta_{\pm\mathfrak{i}}$ is the cosine of the angle between the outgoing particle's momentum $\bm{p}_\pm$ and an incoming particle's momentum $\bm{p}_\mathfrak{i}$, either particle $\mathfrak{a}$ or $\mathfrak{b}$. 

\begin{widetext}
\begin{subequations}
\begin{align}
    \label{eqn: binary gain matrix delta removed}
    G_{\mathfrak{a}\mathfrak{b}\rightarrow\mathfrak{c}\mathfrak{d},ijklmnopq}&=\int_{P_{ijk}}\int_{P_{lmn}}\int_{P_{opq}}\tilde{G}_{\mathfrak{a}\mathfrak{b}\rightarrow\mathfrak{c}\mathfrak{d}}\frac{\mathrm{d}^3p_\mathfrak{b}}{(p_\mathfrak{b})^2\Delta p_{\mathfrak{b},o}\Delta u_{\mathfrak{b},p}\Delta \phi_{\mathfrak{b},q}}\frac{\mathrm{d}^3p_\mathfrak{a}}{(p_\mathfrak{a})^2\Delta p_{\mathfrak{a},l}\Delta u_{\mathfrak{a},m}\Delta \phi_{\mathfrak{a},n}}\d u_\mathfrak{c}\d\phi_\mathfrak{c},
    \\ 
    \label{eqn: binary gain term delta removed}
    \tilde{G}_{\mathfrak{a}\mathfrak{b}\rightarrow\mathfrak{c}\mathfrak{d}}&=\sum_{\pm}\frac{p_\pm^2}{\left|p^0_\mathfrak{a}p_\pm-p^0_\pm p_\mathfrak{a}\cos\Theta_{\pm\mathfrak{a}}+p^0_\mathfrak{b}p_\pm-p^0_\pm p_\mathfrak{b}\cos\Theta_{\pm\mathfrak{b}}\right|}\frac{1}{\left(1+\delta_{\mathfrak{ab}}\right)\left(1+\delta_{\mathfrak{cd}}\right)}\frac{\mathcal{F}_{\mathfrak{a}\mathfrak{b}}^2}{\pi p^0_\mathfrak{a}p^0_\mathfrak{b}}\frac{\mathrm{d}\sigma_{\mathfrak{a}\mathfrak{b}\rightarrow\mathfrak{c}\mathfrak{d}}}{\mathrm{d}T},
\end{align}
\end{subequations}
\end{widetext}

This presents the problem discussed in \sref{subsec: importance sampling}, i.e. that the values of $p_\mathfrak{c}$ cannot be directly sampled, even though the spectrum of outgoing particles may have a peak around a specific range of outgoing angles. Without inverting the differential cross section, this peak may be identified by a simple statement\,---\,\textit{particles are more likely to scatter towards the direction of the centre of momentum velocity}. The simplest approach to importance sampling is to sample the angles $u$ and $\phi$ uniformly in the centre of momentum frame. These angles are then de-boosted and rotated into the observer frame to generate the outgoing angles $u_\mathfrak{c}$ and $\phi_\mathfrak{c}$. However there are two complications to this simple approach: first is that scattering may not be uniform in the centre of momentum frame, for example inverse-Compton scattering in the relativistic regime favours scattering beamed along the centre of momentum direction even in the centre of momentum frame, the second is that for massive particles the de-boosting of angles does not follow the standard Doppler formulae\,---\,requiring two different methods for massive and massless particles. 

This work takes a novel approach to solve both problems: use the Doppler boosting formulae but allow the rapidity of the boost to be a free variable which can be scaled up and down to sample different angular regions, i.e. the Doppler boosting formulae give the probability $P(u,\phi,\rho)$ of sampling the outgoing angles for which the rapidity $\rho=\tanh^{-1}{\beta}$, with $\beta=v/c$, of the boost can be used as a scaling parameter. Given a rapidity $\rho$ and a uniformly sampled set of angles $u$ and $\phi$, an observer frame set of angles $u'_\mathfrak{c}$ and $\phi'_\mathfrak{c}$ can be produced using the Doppler formulae:
\begin{equation}\label{eqn: doppler angles}
    u'_\mathfrak{c}=\frac{u+\tanh{\rho}}{1+u\tanh{\rho}}, \quad \phi'_\mathfrak{c}=\phi.
\end{equation}
The associated probability of sampling these primed angles is neatly given by the Doppler boosting factor for the solid angle element:
\begin{equation}\label{eqn: prob doppler angles}
    P(u,\phi,\rho)=\left(\cosh(\rho)+u\sinh(\rho)\right)^2
\end{equation}
The primed angles are measured with respect to the direction of the centre of momentum frame $\boldsymbol{\beta}^{*}$ in the observer frame:
\begin{equation}\label{eqn: com velocty}
    \beta^{*i}=\frac{p_\mathfrak{a}^i+p_\mathfrak{b}^i}{p_\mathfrak{a}^0+p_\mathfrak{b}^0},~~\cos\theta^*=\frac{\beta^{*3}}{\left|\boldsymbol{\beta}^*\right|},~~ \tan\phi^*=\frac{\beta^{*2}}{\beta^{*1}}.
\end{equation} 
The observer frame angles $u_\mathfrak{c}$ and $\phi_\mathfrak{c}$ are then obtained by rotating the primed angles $u'_\mathfrak{c}$ and $\phi'_\mathfrak{c}$ by the centre of momentum vector angles $\theta^*$ and $\phi^*$. 

Further, for massive particles not all values of $u'_\mathfrak{c}$ are necessarily accessible, with the maximum value given by:
\begin{equation}\label{eqn: obs limit}
    u'_{\mathfrak{c},\text{max}} = \sqrt{1-\left(\frac{p^*}{m_\mathfrak{c}\sinh \rho^*}\right)^2},
\end{equation}
where $p^*$ and $\rho^*$ are the momentum of particle in the centre of momentum frame and rapidity of the centre of momentum frame respectively (see \aref{app: angle limits} for more detail). The limit given by \eref{eqn: obs limit} only exists if $p^*/m_\mathfrak{c}\sinh \rho^*\leq 1$. If it does exist, then sampling of $u'_\mathfrak{c}$ are weighted by a rapidity $\tilde{\rho}$ (\eref{eqn: tilde rap}) such that in the frame moving with that rapidity the angle $u$ corresponding to $u'_\mathfrak{c}$ is equal to zero, i.e. half of the samples lie within this observer frame angular limit. 

Using this weighted sampling approach, the Monte-Carlo estimate for the gain array elements (\esref{eqn: binary gain matrix delta removed} and \eqref{eqn: binary gain term delta removed}) is then given by:
\begin{equation}
\begin{split}\label{eqn: binary gain matrix MC}
    &G_{\mathfrak{a}\mathfrak{b}\rightarrow\mathfrak{c}\mathfrak{d},ijklmnopq}=\frac{\Delta u_{\mathfrak{c},j}\Delta\phi_{\mathfrak{c},k}}{N_\text{loss}N_\text{gain}}
    \\
    &\quad\quad\times\sum_{\alpha=1}^{N_\text{loss}}\sum_{\beta=1}^{N_\text{gain}}\frac{\tilde{G}_{\mathfrak{a}\mathfrak{b}\rightarrow\mathfrak{c}\mathfrak{d}}(\{\bm{p}_\mathfrak{a},\bm{p}_\mathfrak{b}\}_\alpha,\{u,\phi\}_\beta)}{P(u,\phi,\rho)}.
\end{split}
\end{equation}

The Monte-Carlo integration procedure for binary interactions as executed in \texttt{DiplodocusCollisions.jl} is as follows:
\vspace{-5pt}
\begin{algorithm}
\DontPrintSemicolon
$\textbf{def}~s$\;
\For{$N_\text{loss}$}{
    randomly sample $p_\mathfrak{a},u_\mathfrak{a},\phi_\mathfrak{a},p_\mathfrak{b},u_\mathfrak{b},\phi_\mathfrak{b}$\;
    locate loss array element corresponding to $p_\mathfrak{a},u_\mathfrak{a},\phi_\mathfrak{a},$ $p_\mathfrak{b},u_\mathfrak{b},\phi_\mathfrak{b}$\;
    \text{evaluate $L_{\mathfrak{c}\mathfrak{d}\leftarrow\mathfrak{a}\mathfrak{b}}$ using \eref{eqn: binary loss and gain terms}}\;
    update loss array elements using \eref{eqn: binary loss matrix MC}\;
    \For{$N_\text{gain}$}{
        randomly sample $u$ and $\phi$\;
        \eIf{$p^*/m_\mathfrak{c}\sinh \rho^*\leq 1$}{
          $\rho = s\tilde{\rho}$\;
          }{
          $\rho = s\rho^{*}$\;
        }
        evaluate $u_\mathfrak{c}$ and $\phi_\mathfrak{c}$ using the boosting and rotations provided by \esref{eqn: doppler angles} and \eqref{eqn: com velocty}\;
        evaluate $p_\mathfrak{c}$ using energy conservation\;
        locate gain array element corresponding to $p_\mathfrak{a},u_\mathfrak{a},\phi_\mathfrak{a},p_\mathfrak{b},u_\mathfrak{b},\phi_\mathfrak{b}$, and $p_\mathfrak{c},u_\mathfrak{c},\phi_\mathfrak{c}$\;
        evaluate $P(u,\phi,\rho)$ using \eref{eqn: prob doppler angles}\;
        evaluate $\tilde{G}_{\mathfrak{a}\mathfrak{b}\rightarrow\mathfrak{c}\mathfrak{d}}$ using \eref{eqn: binary gain matrix delta removed}\;
        update gain array elements using \eref{eqn: binary gain matrix MC}\;
    }
}
evaluate weighted average of gain array elements using \eref{eqn: weighting scheme} (if previous integral estimate exists)\;
\caption{Binary Interaction Monte-Carlo}
\end{algorithm}
\vspace{-5pt}

\begin{figure}[!ht]
    \centering
    \includegraphics{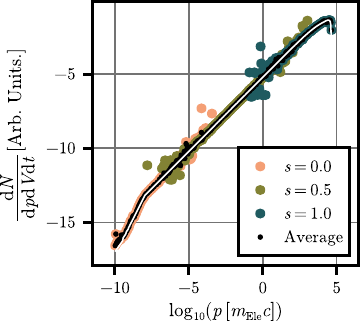}
    \caption{\label{fig: good bad ugly}Isotropic emission spectrum for inverse-Compton scattering of photons with momenta $p=\qty{e-8}\,[m_ec]$ by electrons with momenta $p=\qty{e6}\,[m_ec]$, generated using the methods discussed in \ssref{subsec: importance sampling} and \ref{subsec:weighting}, with the differential cross section given in \aref{app: diff x sections}. The scale factor $s$ adjusts which angles of the outgoing photon are sampled and therefore affect the area of the dependent photon momentum that is sampled. The total integral estimate (black dots) is then calculated using a weighted average (\eref{eqn: weighted average}) and closely matches the approximate spectrum given by \citet{sarkarDissectingComptonScattering2019} (white line).}
\end{figure}

An example of this integration process for inverse-Compton scattering can be seen in \fref{fig: good bad ugly}; different scaling $s$ of the rapidity sample different regions of the outgoing, up-scattered photons momentum (the dependent variable). The spectrum (collision rate) of these outgoing photons is then given by a weighted average of these different integration estimates, closely matching the approximated spectrum given by \citet{sarkarDissectingComptonScattering2019}, to within the noise of the Monte-Carlo sampling, which reduces with the number of points sampled.

\subsubsection{Monte-Carlo Noise Correction for Binary Interactions}{\label{subsubsec: binary correction}}

The noise introduced by Monte-Carlo sampling of binary-collision matrices, if left unchecked, may lead to poor conservation of particle number density and energy density. Each loss array element describes directly the rate of loss of particle number density from a sub-domain in momentum space but also indirectly the rate of loss of energy density by multiplication with the mean energy of that sub-domain. The net rate of gain of particles from that incoming sub-domain in momentum is then given by the sum of associated gain array elements, and likewise the net energy gain can be calculated by the sum of those elements multiplied by the mean energy of the sub-domain that each element corresponds to.

By comparing these loss and gain rates, introduced by imperfect Monte-Carlo integration, a corrective scaling is applied to the gain array elements such that both properties are conserved to machine precision, while maintaining the spectral shape of outgoing particles from an interaction.

Plots describing the conservation of number and energy density for the test cases presented in \sref{sec: test cases} can be found in \aref{app: num eng den}.

\subsection{Emissive Interactions}\label{sec: emissive interactions}
The gain array elements for a particle of type-$\mathfrak{c}$ emitted by a particle of type-$\mathfrak{a}$ in the emissive interaction $\mathfrak{a}\rightarrow\mathfrak{b}\mathfrak{c}$ is given in general by Eq.~(A12) of \citetalias{EverettCotter_2025}: 
\begin{subequations}
\begin{align}
    \label{eqn: emissive gain matrix}
    \begin{split}
    G_{\mathfrak{a}\rightarrow\mathfrak{b}\mathfrak{c},ijklmn}&=\int_{P_{ijk}}\int_{P_{lmn}}G_{\mathfrak{a}\rightarrow\mathfrak{b}\mathfrak{c}}
    \\
    &\quad\times\frac{\mathrm{d}^3p_\mathfrak{a}}{(p_\mathfrak{a})^2\Delta p_{\mathfrak{a},l}\Delta u_{\mathfrak{a},m}\Delta \phi_{\mathfrak{a},n}}\frac{\mathrm{d}^3p_\mathfrak{c}}{(p_\mathfrak{c})^2}, 
    \end{split}
    \\
    \label{eqn: emissive gain term}
    G_{\mathfrak{a}\rightarrow\mathfrak{b}\mathfrak{c}} &= (p_\mathfrak{c})^2\frac{\mathrm{d}N_{\mathfrak{a}\rightarrow\mathfrak{b}\mathfrak{c}}}{\mathrm{d}x^0\mathrm{d}^3p_\mathfrak{c}},  
\end{align}
\end{subequations}
where $\mathrm{d}N_{\mathfrak{a}\rightarrow\mathfrak{b}\mathfrak{c}}/\mathrm{d}x^0\mathrm{d}^3p_\mathfrak{c}$ is the emissivity of the interaction. 

Assuming the emissivity is known (for the example of synchrotron emissions see \aref{app: synchrotron}) this can be integrated using the same Monte-Carlo sampling techniques discussed in \sref{subsec: binary collisions}, without the complication of any importance sampling as there is no dependent variable:
\begin{equation}
\begin{split}\label{eqn: emissive gain matrix MC}
    G_{\mathfrak{a}\rightarrow\mathfrak{b}\mathfrak{c},ijklmn}=&\frac{\Delta p_{\mathfrak{c},i}\Delta u_{\mathfrak{c}_j}\Delta\phi_{\mathfrak{c},k}}{N_\text{loss}N_\text{gain}}
    \\
    &\times\sum_{\alpha=1}^{N_\text{loss}}\sum_{\beta=1}^{N_\text{gain}}G_{\mathfrak{a}\rightarrow\mathfrak{b}\mathfrak{c}}(\{\bm{p}_\mathfrak{a}\}_\alpha,\{\bm{p}_\mathfrak{c}\}_\beta).
\end{split}
\end{equation}

The Monte-Carlo integration procedure for emissive interactions is executed in \texttt{DiplodocusCollisions.jl} as follows:
\vspace{-5pt}
\begin{algorithm}
\DontPrintSemicolon
\For{$N_\text{loss}$}{
    randomly sample $p_\mathfrak{a},u_\mathfrak{a},\phi_\mathfrak{a}$\;
    \For{$N_\text{gain}$}{
      randomly sample $p_\mathfrak{c},u_\mathfrak{c},\phi_\mathfrak{c}$\;
      locate gain array element corresponding to $p_\mathfrak{a},u_\mathfrak{a},\phi_\mathfrak{a}$, and $p_\mathfrak{c},u_\mathfrak{c},\phi_\mathfrak{c}$\;
      evaluate ${G}_{\mathfrak{a}\rightarrow\mathfrak{b}\mathfrak{c}}$ using \eref{eqn: emissive gain term}\;
      update gain array elements using \eref{eqn: emissive gain matrix MC}\;
    }
}
\caption{Emissive Interaction Monte-Carlo}
\end{algorithm}
\vspace{-5pt}

%% file: Sections/transport.tex
\section{Transport}\label{sec:transport}

The role of \texttt{DiplodocusTransport.jl} is to evaluate the transport equation for the particle distribution function, in DIP form, as described by Eqs.~(C1) to (C13) of \citetalias{EverettCotter_2025}. These equations involve the evaluation of the fluxes $\mathcal{A},\mathcal{B},\mathcal{C},\mathcal{D},\mathcal{I},\mathcal{J},$ and $\mathcal{K}$ on the sub-domain boundaries of the phase space coordinates $t,x,y,z,p,u,$ and $\phi$ respectively. As shown in \citetalias{EverettCotter_2025}, these flux terms are independent of the distribution function in DIP form. Therefore, the spacetime fluxes $\mathcal{A},\mathcal{B},\mathcal{C},\mathcal{D}$ only depend on the spacetime metric $g_{\alpha\beta}$ and coordinate transform $e_a^{~\alpha}$ between the coordinate basis of spacetime and local orthonormal basis of momentum space must be defined, whereas the momentum fluxes $\mathcal{I},\mathcal{J},\mathcal{K}$, require the additional knowledge of the Ricci rotation coefficients (connection coefficients in a local orthonormal basis) $\Gamma^a_{~bc}$ and external forces $F^{a}$. With those terms provided, the evaluation of the fluxes on sub-domain boundaries of phase space can be performed analytically and tabulated as a set of seven dimensional arrays.

\subsection{Numerical Scheme}\label{subsec: scheme}

\begin{figure}[!ht]
    \centering
    \includegraphics[]{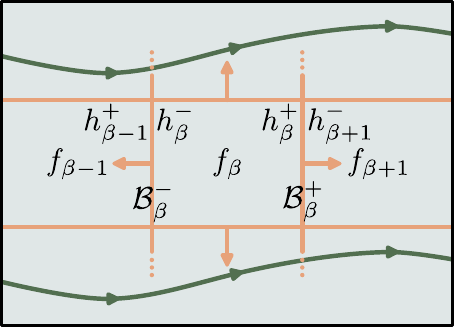}
    \caption{Locations of the values $h^\pm$ and the fluxes $\mathcal{B}^\pm$ with respect to boundaries of phase-space sub-domains (grid cells) in the $\beta$ coordinate direction.}
    \label{fig: num scheme example}
\end{figure}

Within the DIPLODOCUS framework the exact nature of the numerical scheme described by Eq.~(C1) of \citetalias{EverettCotter_2025} depends on $h^\pm$, the values of the boxcar function, used to define DIP (see \eref{eqn: DIP Boxcar}), at the boundaries of phase-space sub-domains. For this work, an \textit{upwind} scheme is used for all fluxes across boundaries of physical-space and momentum-space sub-domains, i.e. $\mathcal{B},\mathcal{C},\mathcal{D},\mathcal{I},\mathcal{J},\text{ and } \mathcal{K}$ fluxes, as it was found to be most stable. An upwind scheme dictates that the evolution of the particle distribution function within a sub-domain of phase space depends only on the distribution in sub-domains upwind with respect to the phase flow. For example, consider \fref{fig: num scheme example}, the phase flow is along the direction of increasing $\beta$ coordinate, from left to right, therefore the values of $\mathcal{B}^-_\beta$ and $\mathcal{B}^+_\beta$ are negative and positive respectively\footnote{The direction of fluxes is taken to be the same as the outward pointing normal to the grid sub-domain boundaries.}. For an upwind scheme the rate of transfer of particles is dependent on their distribution upstream with respect to the flow, to achieve this the values of $h^{+}_{\beta-1}$ and $h^{+}_\beta$ are taken to be unity and $h^{-}_{\beta}$ and $h^{-}_{\beta-1}$ are taken to be zero. This process is then repeated identically for all boundaries across physical space and momentum space. As the values and signs of the fluxes are independent of the distribution function, the determination of the value of $h^{\pm}$ may be computed in advance.

For the flux in the time direction $\mathcal{A}$, there are two obvious choices: \textit{upwind} and \textit{downwind} (more commonly explicit (forwards) and implicit (backwards) Euler methods) where downwind is the opposite of upwind such that evolution is dependent on the distribution of particles downstream in time. The upwind scheme in time is known to be susceptible to numerical instability if the Courant number for the system $Cr\propto \Delta t$ becomes greater than one (known as a Courant-Friedrichs-Lewy (CFL) condition \citep{CourantEtAl_1928}), which may occur if the time step (distance between boundaries along the time coordinate) is sufficiently large. The downwind scheme on the other hand, is known to be numerically stable and particularly useful for stiff equations where there is a large separation of interaction timescales \citep{hairerSolvingOrdinaryDifferential2009,hairerSolvingOrdinaryDifferential2010}\,---\,making it ideal for particle transport through phase space with a wide range of interactions and forces. However, in practice this involves a matrix factorisation and division which are prone to numerical instability through loss of precision, as was found here. Hence the time stepping scheme for this paper is taken to be upwind while a more stable downwind scheme is under development.

\subsection{Array Flattening}\label{subsec: array flat}
The evaluation of collision integrals (Section \ref{sec:collisions}) generates arrays of high dimension (up to nine for binary interactions and six for emissive) which act only on the momentum space of the system. Similarly the evaluation of fluxes generates seven dimensional arrays that act on all dimensions of phase space. As the dimension of phase space is also seven (one time, three space and three momentum space), the normalised distribution functions $f_{\alpha\beta\gamma\delta ijk}$ that describe the state of the system in phase space are also described by seven dimensional arrays. For computation, the dimension of time is not saved at every time step, and all forces are assumed to be independent of time, therefore, in practice the state of the system and fluxes are six dimensional arrays.

The evaluation of the transport equation (\citetalias{EverettCotter_2025}, Eq.~(C1)) relies on multiplication of these high dimensional arrays. For general hardware, the multiplication of multi-dimensional arrays is not as optimised as matrix-matrix and matrix-vector multiplication. Therefore the distribution functions $f_{\mathfrak{a},\alpha\beta\gamma\delta ijk}$ for all species $\mathfrak{a}$, are merged into a single dimensional column vector $f_{a}$ where 
\begin{equation}
\begin{split}
    a &= (k_\mathfrak{a}-1)N_{p,\mathfrak{a}}N_{u,\mathfrak{a}}+(j_\mathfrak{a}-1)N_{p,\mathfrak{a}}+i_\mathfrak{a}+O_{\mathfrak{a}}
    \\
    &\quad+(\beta-1)N_yN_z+(\gamma-1)N_z+\delta-1,
\end{split}
\end{equation}
where $N_{p,\mathfrak{a}}, N_{u,\mathfrak{a}}$ and $N_{\phi,\mathfrak{a}}$ are the number of momentum space grid cells for species $\mathfrak{a}$ and $O_{\mathfrak{a}}$ is the species offset in the column vector given by \begin{equation}
    O_\mathfrak{a} = \sum_{\mathfrak{a}=1}^{\mathfrak{a}-1} N_{p,\mathfrak{a}}N_{u,\mathfrak{a}}N_{\phi,\mathfrak{a}}.
\end{equation}

The transport equation (\citetalias{EverettCotter_2025}, Eq.~(C1)) can similarly be re-written (according to the numerical scheme described in \sref{subsec: scheme}) in a flattened form:
\begin{equation}\label{eqn: flat transport w.o. coll}
    \mathcal{A}^+_{ab}\Delta f^{}_{b} + \mathcal{F}^{}_{ab}f^{}_{b} = \text{collisions terms},  
\end{equation}
where repeated indices imply matrix multiplication, and $\Delta f_{a}$ is the change in $f_a$ between time steps $\alpha$ and $\alpha+1$. The flux matrix $\mathcal{A}^+_{ab}$ is the result of flattening the higher dimensional flux array $\mathcal{A}^+$, likewise, the flux matrix
\begin{equation}
    \mathcal{F}_{ab}=\sum_{\pm}\mathcal{A}^\pm_{ab}+\mathcal{B}^\pm_{ab}+\mathcal{C}^\pm_{ab}+\mathcal{D}^\pm_{ab}+\mathcal{I}^\pm_{ab}+\mathcal{J}^\pm_{ab}+\mathcal{K}^\pm_{ab},
\end{equation}
is the sum of all the flattened higher dimensional flux arrays.

\subsubsection{Collision Terms}\label{subsubsec: flattening collision terms}
Collision terms are independent of spacetime and only act on the momentum space of the particles. For emissive interactions $\mathfrak{a}\rightarrow\mathfrak{bc}$, the rate is dependent only on a single particle's momentum space and therefore can be written as a two dimensional matrix $\mathcal{M}_{\text{Emi},\tilde{a}\tilde{b}}$ akin to the phase space fluxes, however, unlike these fluxes, the size of the emission matrix is smaller being that it is independent of spacetime, with 
\begin{equation}
\begin{split}
    \tilde{a} &= (k_\mathfrak{a}-1)N_{p,\mathfrak{a}}N_{u,\mathfrak{a}}+(j_\mathfrak{a}-1)N_{p,\mathfrak{a}}+i_\mathfrak{a}+O_{\mathfrak{a}},
    \\
    \tilde{b} &= (k_\mathfrak{b}-1)N_{p,\mathfrak{b}}N_{u,\mathfrak{b}}+(j_\mathfrak{b}-1)N_{p,\mathfrak{b}}+i_\mathfrak{b}+O_{\mathfrak{b}},
\end{split}
\end{equation} 
for the emission of a particle of type-$\mathfrak{a}$ from a particle of type-$\mathfrak{b}$.

For binary interactions $\mathfrak{ab}\leftrightharpoons\mathfrak{cd}$, the rate is dependent on in momenta of the two incoming particles, as such flattening produces a three dimensional array $\mathcal{M}_{\text{Bin},\tilde{a}\tilde{b}\tilde{c}}$, requiring multiplication with the state vector twice. In practice this is achieved as repeated matrix-vector multiplication by first flattening $\mathcal{M}_{\text{Bin},\tilde{a}\tilde{b}\tilde{c}}$ to a two dimensional matrix to perform the multiplication with the first state vector and then reshaping the resulting vector to a matrix for multiplication with the state vector for a second time. 

These collision terms can then be included in \eref{eqn: flat transport w.o. coll}:
\begin{equation}\label{eqn: flat transport w. coll}
\begin{split}
    \mathcal{A}^+_{ab}\Delta f^{}_{b} + \mathcal{F}^{}_{ab}f^{}_{b} = \sum_{\beta,\gamma,\delta\in b,c} & \left( \mathcal{M}_{\text{Emi},\tilde{a}\tilde{b}}f^{}_{b} + \right. 
    \\
    &\left. \quad\mathcal{M}_{\text{Bin},\tilde{a}\tilde{b}\tilde{c}}f^{}_{b}f^{}_{c}\right)\mathcal{V}_{\beta\gamma\delta},
\end{split}
\end{equation}
where $\mathcal{V}_{\beta\gamma\delta}$ is the volume element for each spatial sub-domain (\citetalias{EverettCotter_2025}, Eq.~(C2)), resulting in the transport equation implemented directly into \texttt{DiplodocusTransport.jl}. Given the assumption that all forcing terms are independent of time, the matrix $\mathcal{A}^+_{ab}$ is diagonal, hence direct element-wise division is used rather than matrix factorisation to generate the change in state vector $\Delta f_{b}$.

%% file: Sections/test_cases.tex
\section{Test Cases}\label{sec: test cases}

This section is dedicated to a series of test cases to demonstrate the qualitative and quantitative ability of the \texttt{Diplodocus.jl} code and any limitations that may be the subject of improvement in later works.

As discussed in \sref{sec:intro} this paper will only consider tests of micro-scale physical effects, neglecting the spatial advection that will be the focus of \citetalias{PaperIII}. As such all particle distributions in this work are taken to be \textit{homogenous in space}. 

This homogenous space is taken to be flat (Minkowski) with the choice of coordinate basis depending on the test case (\aref{app: coords}). Each coordinate basis has its own definition for the local orthonormal tetrad $e_a^{~\alpha}$ and Ricci rotation coefficients $\Gamma^{a}_{~bc}$ that are required to generate the fluxes described in \citetalias{EverettCotter_2025} and \sref{sec:transport}.

As of this work, a range of forces and interactions relevant for the study of astrophysical jets that consist of \textit{leptonic} particles have been implemented in \texttt{Diplodocus.jl}. These include: Lorentz forces, radiation reaction and synchrotron emission (\aref{app: rad react sync}), (inverse-)Compton scattering (\aref{app: compt x-sec}), electron-positron annihilation to two photons (\aref{app: pair anni x-sec}), and electron-positron pair production via the annihilation of two photons (\aref{app: photon anni x-sec}).    

The evaluation of all collision arrays and transport equations for this section were performed on a single desktop system supporting 20-core Intel i7-12700 and \qty{64}{GB} of RAM. Evaluation of collision arrays via \texttt{DiplodocusCollisions.jl} is conducted with 64-bit precision and support multi-threading to increase the number of Monte-Carlo samples. Evaluation of transport equations via \texttt{DiplodocusTransport.jl} was conducted with 32-bit precision and is currently single-threaded. 

The momentum-space grids used for the test cases presented in this section were chosen as a compromise between accuracy and memory intensity, bounded by the limited RAM within the system used to perform these tests. The most intensive component is the collision matrix $\mathcal{M}_{\text{Bin},\tilde{a}\tilde{b}\tilde{c}}$ found within the transport equation \eref{eqn: flat transport w. coll}, which scales with the total number of momentum-space grid cells for all particle species cubed, as is dependent on the initial state of both incoming particles in a binary interaction and the state of the outgoing particle.

\input{Sections/Test_Cases/ideal_gas.tex}

\input{Sections/Test_Cases/ExB}

\input{Sections/Test_Cases/radiation_reaction.tex}

\input{Sections/Test_Cases/synchrotron.tex}

%% file: Sections/Test_Cases/ideal_gas.tex
\begin{figure*}[!ht]
    \centering
    \vspace{11pt}
    \includegraphics[]{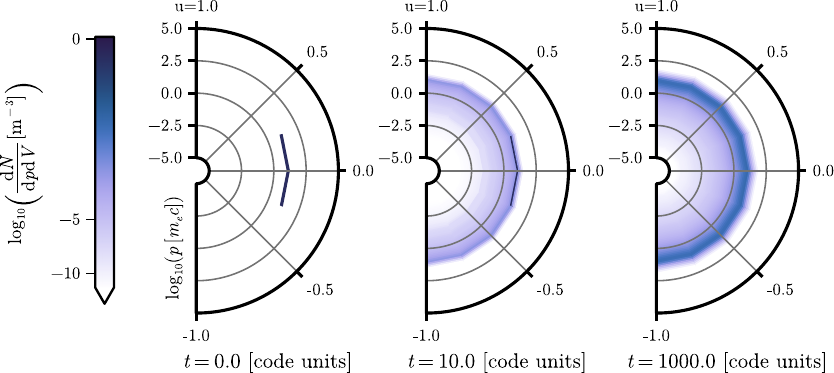}
    \caption{\label{fig: hard sphere p and u dis}Time evolution of the hard sphere distribution in momentum space undergoing elastic collisions. The distribution is presented in polar coordinates, with $\theta=\arccos u$, $r=\log_{10}p$, with the colour scale indicating the magnitude of the distribution.}
\end{figure*}

\subsection{\label{subsec: hard spheres}Elastically Colliding Spheres}

To demonstrate that \texttt{Diplodocus.jl} is functioning at a basic level, the first test cases considered is that of the interaction between hard spheres. ``Hard spheres" here refer to a population of perfect spheres, all of common radius ($R_\circ=\sqrt{\sigma_T/\pi}$) and mass ($m_\circ=1836\,m_e$), interacting via elastic, binary collisions (the cross section for which is given in \aref{app: hard sphere x-sec}).

If this population has an initially non-thermal distribution, via the effect of the interaction between particles, the population is expected to tend towards a thermal (Maxwell-J\"uttner) distribution, likewise, if the population is initially anisotropic (but with zero bulk velocity) it is expected to tend towards isotropy in the observer frame. To examine whether DIPLODOCUS can reproduce this behaviour the collision matrix was evaluated for a momentum range of $p=10^{-5}\,m_ec$ to $10^{4}\,m_ec$ with a momentum space grid of $N_p\times N_u\times N_\phi=72\times8\times1$ ($p$ bins are uniformly spaced in $\log_{10}p$ and $N_\phi=1$ corresponds to the assumption that particles are distributed axisymmetrically in momentum space). The spatial coordinates are taken to be Cartesian (as described in \aref{app: cart coords}) and the initial conditions of this population were taken to be a ``delta function" in momentum space, i.e. a single bin, which ranges from $p=10\,m_ec$ to $13.3\,m_ec$, and a polar angle range from $u=-0.25$ to $0.25$.

\fref{fig: hard sphere p and u dis} shows the evolution of this system over time as a function of $p$ and $u$, with the first panel showing the initial conditions at $t=0$. Over several characteristic timescales, the spheres spread out in momentum space as a result of the elastic collisions.

\begin{figure}[!ht]
    \centering
    \vspace{5pt}
    \includegraphics[]{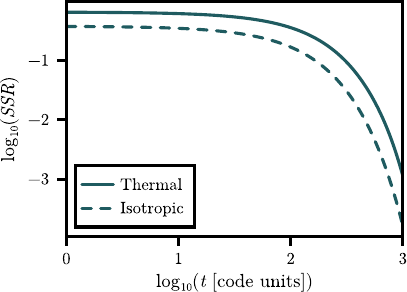}
    \caption{\label{fig: hard sphere T and I}Time evolution of the Sum of Squared Residuals (SSR) between the particle distribution and the expected isotropic (dashed line) and thermal (solid line) distributions.}
\end{figure}

\begin{figure}[!ht]
    \centering
    \includegraphics[]{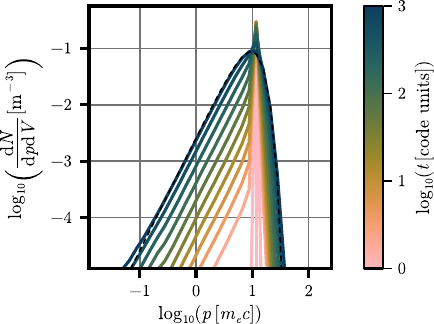}
    \caption{\label{fig: hard sphere p dis}Time evolution of the angle averaged momentum distribution of hard spheres undergoing elastic collisions. Additionally plotted (dashed black line) is the expected thermal distribution.}
\end{figure}

\fsref{fig: hard sphere T and I} and \ref{fig: hard sphere p dis}, show that this spreading out tends towards the expected thermal and isotropic distribution. However, as can be observed in \fref{fig: hard sphere p dis}, the expected thermal distribution is marginally \textit{over-shot} at momenta either side of the distributions peak. This effect is attributed to numerical diffusion as a result of the finite momentum bin widths. 

As particles are neither created nor destroyed via the elastic collision of hard spheres both number and energy density are expected to be conserved, is demonstrated to machine precision in \fref{fig: hard sphere frac num den frac eng den} of \aref{app: hard sphere num eng den}.

%% file: Sections/Test_Cases/ExB.tex
\subsection{Electron Gyration and Drift}\label{subsec: ExB} 

\subsubsection{Electron Gyration}\label{subsubsec: ExB gyration}

The motion of charged particles under the influence of a Lorentz force induced by a uniform electromagnetic field is well known \citep[see, for example,][]{LandauLifshits_1975}. If only a uniform magnetic field is present, non-relativistic, charged particles gyrate about this field with a gyro-period $t_\text{gyro}=2\pi m/qB$, where $m$ is the mass of the particle, $q$ its charge and $B$ the magnetic field strength. 

It is possible to simulate such circular motion with \texttt{Diplodocus.jl} using a Cartesian grid (\aref{app: cart coords}) with a magnetic field of strength $B=m\sigma_\text{T}c/q$ aligned to the $z$-axis. As such the gyration is in the $xy$-plane and $t_\text{gyro}=2\pi t\,[\text{code units}]$. The momentum-space domain was taken to be $p=[1.1\times10^{-3},10^{-2}]\,m_ec$, $u=[-10^{-3},10^{-3}]$ and $\phi=[0,2\pi]$, with $N_p\times N_u\times N_\phi=160\times1\times128$ ($p$ bins are uniformly spaced in $p$) and initially electrons are placed with momentum around $5\times10^{-3}\,[m_ec]$ at an azimuthal angle of $\pi$ in momentum-space. As the Lorentz force only acts in the azimuthal direction in momentum space, the Courant number may be calculated as $Cr=u_\phi \Delta t/\Delta \phi$, where $u_\phi$ is the constant velocity along the azimuthal direction $u_\phi=2\pi/t_\text{gyro}$. Therefore given $\Delta \phi$ is defined by the momentum-space grid, $\Delta t$ was chosen to achieve $Cr\approx1$ for stability and reduced numerical diffusion. 

\begin{figure}[!ht]
    \centering
    \includegraphics[]{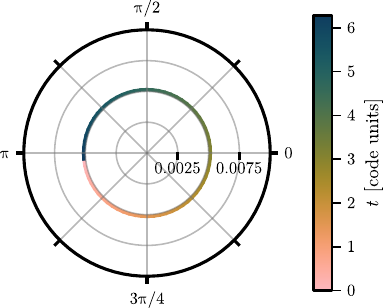}
    \caption{\label{fig: ExB no E p and phi dis}Time evolution of the momentum $p$ (radial coordinate, in units of $m_ec$) and azimuthal angle $\phi$ for a single gyration of a distribution of electrons orbiting a uniform magnetic field.}
\end{figure}

The anti-clockwise gyration of electrons can be observed in \fref{fig: ExB no E p and phi dis}, matching the expected theoretical behaviour of completing a single gyration in a gyro-period, with no diffusion in both the momentum and azimuthal angle directions.

\subsubsection{Electron Drift}\label{subsubsec: ExB drift}
If an electric field of strength $E$ is introduced, perpendicular to the magnetic field, it will accelerate charged particles along its direction, increasing their momenta and the magnitude of the Lorentz force generated by the magnetic field, altering their direction of motion. These two effects cause the gyration of charged particles to drift at a constant velocity $\beta_\text{drift}=E/B$ perpendicular to both the electric and magnetic fields \citep{LandauLifshits_1975}.

\begin{figure}[!ht]
    \centering
    \includegraphics[]{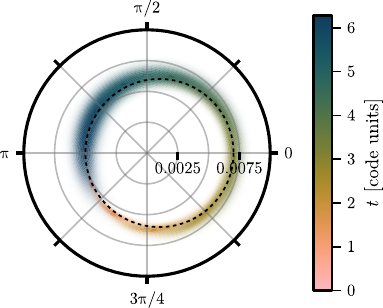}
    \caption{\label{fig: ExB p and phi dis}Time evolution of the momentum $p$ (radial coordinate, in units of $m_ec$) and azimuthal angle $\phi$ distribution of electrons under the action of uniform, perpendicular magnetic and electric fields, compared to the analytical trajectory for a single electron with the same initial conditions (dashed black line). The magnitude of the distribution is indicated the opacity of the colour and the distribution is plotted every 50 time steps for a single gyration.}
\end{figure}

An electric field of strength $E=B/10^3$ aligned to the $y$-axis was added to the setup of \sref{subsubsec: ExB gyration} and the system was evolved in time using the same initial conditions. In momentum space, the drift velocity $\beta_\text{drift}=10^{-3}\approx p_\text{drift}\,[m_ec]$ adds linearly to the circular motion generated by the magnetic field, such that a charged particle's path remains a circle but offset by $\beta_\text{drift}$ in the drift direction. With the electric field in the $y$ direction and magnetic field in the $z$, the drift velocity is expected to be in the $x$ direction ($\phi=0$), as observed in \fref{fig: ExB p and phi dis}.

\begin{figure}[!ht]
    \centering
    \includegraphics[]{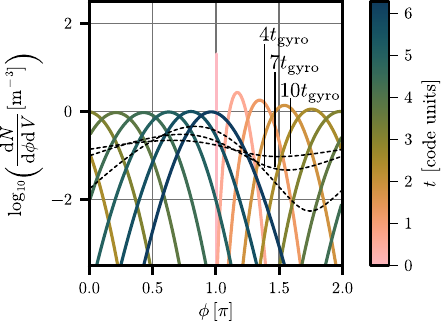}
    \caption{\label{fig: ExB phi dis}Time evolution of the azimuthal angle $\phi$ per gyration for a single gyrations (solid lines) of a distribution of electrons under the action of uniform, perpendicular magnetic and electric fields. Dashed black lines indicate the distribution after $3,7$, and $10$ gyrations.}
\end{figure}

Unlike the simple case of electron gyration, where there is a single, constant flow velocity through momentum space, the action of the electric field causes the flow to depend on the particle's location in momentum space. This introduces multiple speeds at which particles may be advected, and as a result the Courant number can no longer be maintained at unity. Uniform time stepping was chosen to keep $Cr<1$ for the entire simulation, which introduces the numerical diffusion seen in \fsref{fig: ExB p and phi dis} and \ref{fig: ExB phi dis}. This diffusion occurs in both the $p$ and $\phi$ directions, with the distribution tending towards uniformity in $\phi$, as shown in \fref{fig: ExB phi dis} up to 10 gyrations. This effect may be reduced by employing adaptive time stepping and better-optimised momentum-space grids, but that is beyond the scope of this work.

%% file: Sections/Test_Cases/radiation_reaction.tex
\begin{figure*}[!ht]
    \centering
    \includegraphics[]{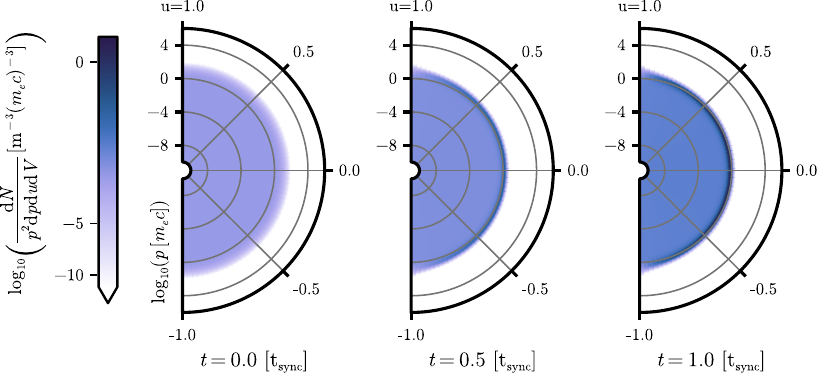}
    \caption{\label{fig: rad react p and u dis}Time evolution of the initially thermal and isotropic distribution of electrons undergoing cooling via a radiation reaction force induced by an axial magnetic field. Electrons with higher momentum cool faster, causing a pile-up at lower momentum and the formation of a ring in momentum space.}
\end{figure*}

\subsection{Radiation Reaction}\label{subsec: ele rad raction}

Under the action of a uniform magnetic field, in addition to a Lorentz force, a population of electrons experiences a radiation reaction force (\eref{eqn: rad react force}), which is a non-conservative force, causing the emission of (synchrotron) photons (see \aref{app: rad react sync}) cooling the population of electrons as well as inducing anisotropy towards the magnetic field axis.

\begin{figure}[!ht]
    \centering
    \includegraphics[]{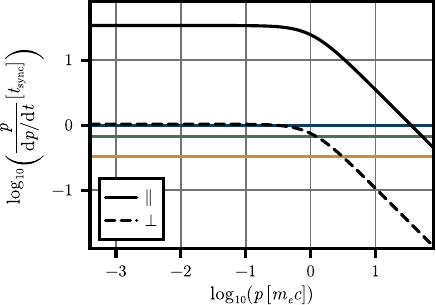}
    \caption{\label{fig: rad react para perp timescale}Timescale for electron momentum loss (\eref{eqn: t sync}), evaluated directly from the $\mathcal{I}$ and $\mathcal{J}$ flux matrices (see \sref{sec:transport}), for the angular bin parallel $\parallel$ and perpendicular $\perp$ to the magnetic field (due to finite angular bin resolution, a distribution of electrons will always have a component of momentum perpendicular to the magnetic field). Coloured horizontal lines correspond to the displayed time steps in \fref{fig: rad react para perp}}
\end{figure}

The timescale at which an electron loses momentum perpendicular to a magnetic field can be calculated from \eref{eqn: rad react power} using the power $P=\mathrm{d}p^0/\mathrm{d}t$, with $p^0$ being the energy in units of momentum, and $\mathrm{d}p/\mathrm{d}p^0=p^0/p$:
\begin{equation}
\begin{split}\label{eqn: t sync}
    t = \frac{p}{\mathrm{d}p/\mathrm{d}t} &= t_\text{sync}\frac{m_ec}{p^0\sin^2\theta},
    \\
    &\approx\begin{cases}
       t_\text{sync}\frac{m_ec}{p\sin^2\theta}, &  \text{for } p\gg m_ec \\
       t_\text{sync}\frac{1}{\sin^2\theta}, &  \text{for } p\ll m_ec\\
    \end{cases}
\end{split}
\end{equation}
where $t_\text{sync}\equiv \mu_0 m_ec/B^2\sigma_\text{T}$  and $\theta$ is the angle between the electron's momentum and the magnetic field direction (see \fref{fig: rad react para perp timescale}).

As a result of \eref{eqn: t sync}, relativistic electrons lose momentum at a faster rate than sub-relativistic electrons potentially causing a \textit{pile-up} (inverted Landau population) in the electron distribution function defined by $\partial f(\bm{p})/\partial p>0$, where $f(\bm{p})\propto \mathrm{d}N/p^2\mathrm{d}p\mathrm{d}u\mathrm{d}V$.

This effect is well studied in \citet{BilbaoSilva_2023,BilbaoEtAl_2024}, which suggest this can best be observed by considering a thermal population of electrons with temperature sufficient such that the peak of their distribution lies around $p=p_\text{th}\approx m_ec$. 

Using \texttt{Diplodocus.jl}, a population of, initially thermally (Maxwell-J\"uttner) and isotropically distributed, electrons was evolved under the influence of a radiation reaction force induced by a magnetic field $B=\qty{e-4}{\tesla}$ directed along the $z$-axis in a cylindrical system (see \aref{app: cyl coords} for details on the coordinates). The initial temperature was taken to be $T=\qty{2.4e10}{\kelvin}$, such that $p_\text{th}=\sqrt{m_ek_BT}=2m_ec$, with a momentum-space domain of $p=[10^{-10},10^{5}]\,m_ec$, $u=[-1,1]$ and $\phi=[0,2\pi]$, with $N_p\times N_u\times N_\phi=480\times65\times1$ ($p$ bins are uniformly spaced in $\log_{10}p$).

\begin{figure}[!ht]
    \centering
    \includegraphics[]{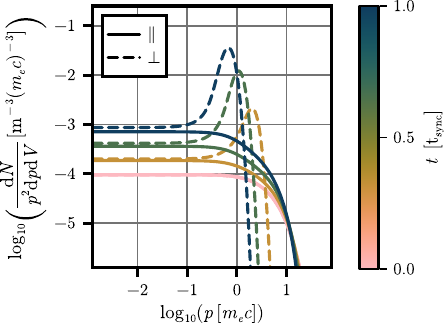}
    \caption{\label{fig: rad react para perp}Time evolution of the components of the electron population parallel $\parallel$ and perpendicular $\perp$ to the magnetic field.}
\end{figure}

\begin{figure*}[!ht]
    \centering
    \vspace{10pt}
    \includegraphics[]{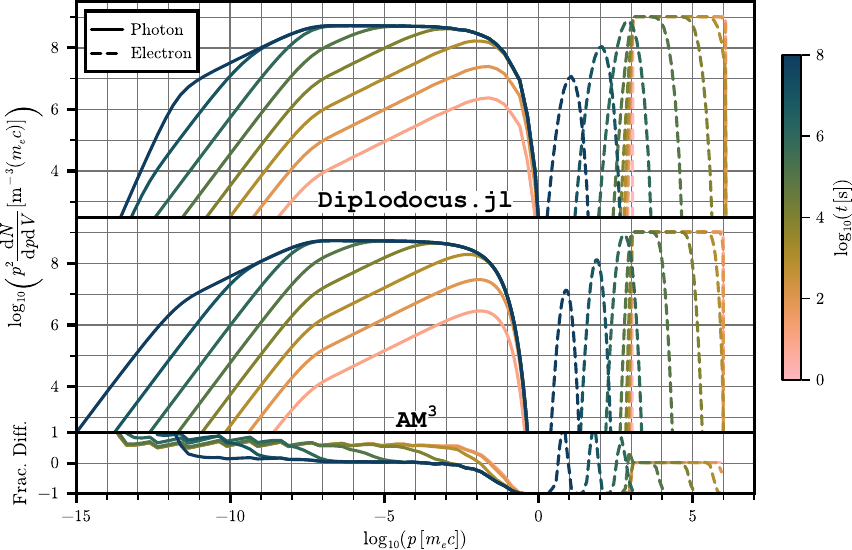}
    \caption{\label{fig: AM3 DIP sync iso}Time evolution of an isotropic system of electrons emitting synchrotron photons due to an isotropically directed magnetic field, as described in \sref{subsec: AM3 sync}, using \texttt{Diplodocus.jl} (top panel) and \texttt{AM\textsuperscript{3}} (middle panel). Fractional difference between the two methods, that is $(f_\text{DIP}-f_\text{AM3})/f_\text{DIP}$, is given in the bottom panel.}
\end{figure*}

\fsref{fig: rad react p and u dis} and \ref{fig: rad react para perp} show the evolution of the electron distribution function over a characteristic synchrotron timescale. Initially the isotropic and thermal electron distribution is flat over momentum with an exponential drop off around $p=m_ec$. However, under the influence of the radiation reaction force, a ring begins to form in the distribution (\fref{fig: rad react p and u dis}) where relativistic electrons begin to pile-up on the sub-relativistic population, the anisotropy of the radiation reaction forces also begins to introduce anisotropies into the electron population, with electrons aligned or anti-aligned to the magnetic field direction ($u=1.0$) cooling slower. The formation of an inverted Landau population can clearly be seen in \fref{fig: rad react para perp} where $\partial f(\bm{p})/\partial p>0$, as cooling relativistic electrons begin to pile up onto the sub-relativistic population.

%% file: Sections/Test_Cases/synchrotron.tex
\subsection{Synchrotron and Synchrotron Self-Compton}\label{sec: ele synchrotron}
\sref{subsec: ele rad raction} demonstrated the DIPLODOCUS framework's ability to evolve electron populations under the influence of a radiation reaction force caused by a uniform magnetic field. This \textit{radiation} reaction force causes the emission of (synchrotron) photons by charged particles. This section contains two direct comparisons between \texttt{Diplodocus.jl} and \texttt{AM\textsuperscript{3}} \citep{KlingerEtAl_2024a} (\ssref{subsec: AM3 sync} and \ref{subsec: AM3 SSC}), to demonstrate the DIPLODOCUS framework's capability to reproduce results obtained by single-zone codes. Furthermore, as \texttt{Diplodocus.jl} is not constrained to the traditional isotropy of single-zone codes, such as \texttt{AM\textsuperscript{3}}, the angular dependence of these tests will then be examined in \sref{subsubsec: ani ssc}.

\subsubsection{\texorpdfstring{$\texttt{AM\textsuperscript{3}}$}{AM3} Comparison: Synchrotron}\label{subsec: AM3 sync} 

As of its current release \texttt{AM\textsuperscript{3}} handles radiation reaction forces and synchrotron emissions assuming an isotropically averaged turbulent magnetic field. This feature is also present as an option within \texttt{DiplodocusTransport.jl}, allowing direct comparison between the two codes. For this comparison, the electron population is initially taken to be isotropic with a power-law distribution in energy ranging from $p=[10^3,10^6]\,m_ec$ with an index of $2$ and an initial number density of $n=\qty{e6}{\meter^{-3}}$. The momentum-space domain of the electrons is taken to be $p=[10^{-3},10^7]\,m_ec$, $u=[-1,1]$ and $\phi=[0,2\pi]$ with $N_p\times N_u\times N_\phi=80\times9\times1$. All $p$ bins are uniformly spaced in $\log_{10}p$ in this section. Note that though the system is isotropic, there is still a range of $u$ bins, as all collision matrices are pre-computed and reusable the range of angular bins will allow anisotropic effects to be examined in \sref{subsubsec: ani ssc}. Initially there are zero photons but the momentum-space domain of photons that are expected to be emitted, due to the presence of a isotropic magnetic field of strength $B=10^{-4}\text{T}$, is taken to be $p=[10^{-15},10^{7}]\,m_ec$, $u=[-1,1]$ and $\phi=[0,2\pi]$ with $N_p\times N_u\times N_\phi=88\times9\times1$. To match \texttt{AM\textsuperscript{3}}, spherical coordinates (\aref{app: sph coords}) are used such that the single-zone is spherical in shape. As space is taken to be homogenous, the system is then evolved without any further injection or escape of particles.

\begin{figure*}[t]
    \centering
    \vspace{10pt}
    \includegraphics[]{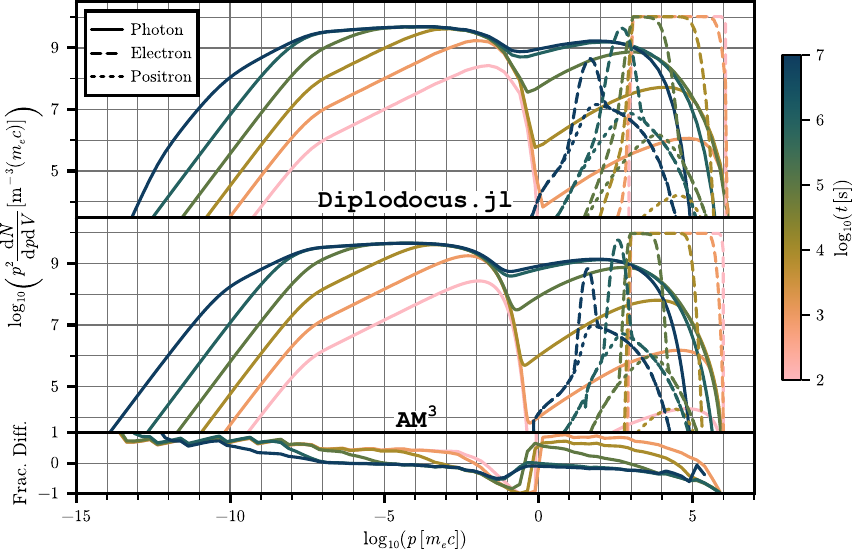}
    \caption{\label{fig: AM3 DIP ssc iso}Time evolution of an isotropic system of electrons emitting synchrotron photons due to an isotropically directed magnetic field,  with those photons then being up-scattered via the process of synchrotron self-Compton and photon-photon pair production generating new electron-positron pairs, as described in \sref{subsec: AM3 SSC}, using \texttt{Diplodocus.jl} (top panel) and \texttt{AM\textsuperscript{3}} (middle panel). Fractional difference between the photon population generated by the two methods is given in the bottom panel.}
\end{figure*}

\fref{fig: AM3 DIP sync iso} shows the evolution of this system using \texttt{Diplodocus.jl} and \texttt{AM\textsuperscript{3}}. The electron population cools, via the radiation reaction force, to lower momenta, building up a photon population via synchrotron emission. The results are almost identical, deviating by less than a factor of $2$ for the majority of the electron and photon distributions, as can be seen with the aid of the fractional difference plotted in the bottom panel of \fref{fig: AM3 DIP sync iso}. However, there are a few regions that show larger differences.

At low photon momentum, the results of \texttt{AM\textsuperscript{3}} and \texttt{Diplodocus.jl} deviate, with \texttt{Diplodocus.jl} having a steeper cut-off. For late time steps, the electron population has cooled to mildly relativistic momenta, where the ultra-relativistic synchrotron approximation used by \texttt{AM\textsuperscript{3}} breaks down. 
\texttt{DiplodocusCollisions.jl} uses a more general synchrotron kernel (\eref{eqn: gain term synchrotron}) that is valid for all electron energies, covering not just the relativistic synchrotron spectrum but also the sub-relativistic cyclotron spectrum and the transition between the two,  causing the deviation in photon spectra below momenta of $\approx10^{-12}\,m_ec$. This deviation may be reduced when  synchrotron self-absorption is included, but this has yet to be implemented within \texttt{Diplodocus.jl}. 

A second deviation to note is slight and refers to differences in the location of the spectral peaks and overall magnitude of the synchrotron spectrum. This is caused by two factors. First, the electron momentum grid used in this test by \texttt{Diplodocus.jl} (8 bins per decade) is coarser than that used by \texttt{AM\textsuperscript{3}} ($\approx$ 23 bins per decade) leading to the cooling of the electron population being more (numerically) diffusive in momenta, as can be seen in \fref{fig: AM3 DIP sync iso}. Secondly, the emission of synchrotron photons and cooling of electron are implemented using two distinct methods. The former is treated via an emission matrix while the latter a force flux term. As the width of bins in momentum space is finite, the Monte-Carlo integration of the emission matrix can cause an over-emission of synchrotron photons near the spectral peak. To ensure energy conservation, a corrective scaling is applied  to the emission matrix, without which energy is only conserved in the limit of infinite grid resolution (see \aref{app: sync and SSC num eng}). However, this corrective term may have the effect of shifting the entire emission spectrum down, leading to the minor underestimation of photon emissions away from the peak (best seen by comparing at the photon spectra at early times in \fref{fig: AM3 DIP sync iso}).

\vspace{20pt}

\subsubsection{\texorpdfstring{$\texttt{AM\textsuperscript{3}}$}{AM3} Comparison: Synchrotron Self-Compton}\label{subsec: AM3 SSC}

Synchrotron self-Compton is a process whereby synchrotron photons emitted by high energy charged particles undergo inverse-Compton scattering from that same population of high energy particles, increasing photon energy. To examine this effect identical initial conditions to those in \sref{subsec: AM3 sync} were chosen for the electron population except that its number density was increased to $n=\qty{e7}{\meter^{-3}}$. Momentum-space grids were left the same such that the synchrotron collision matrix that was pre-computed for \sref{subsec: AM3 sync} could be reused (as is the advantage of pre-computing these matrices) and in addition collision matrices were generated for Compton scattering and photon-photon pair production using the process described in \sref{subsec: binary collisions} and the cross sections found in \asref{app: compt x-sec} and \ref{app: photon anni x-sec}. As photons are up-scattered, the interaction between any two photons may have sufficient centre of mass energy to produce an electron-positron pair, which may replenish the cooling leptonic population\footnote{The counter effect of electron-positron annihilation to a pair of photons is implemented in \texttt{Diplodocus.jl}, with cross sections found in \aref{app: pair anni x-sec}, but was not included in this test as it is not currently implemented in \texttt{AM\textsuperscript{3}}}. The rate of this effect scales with square of the photon population. The initial electron number density was therefore increased (and, consequently, the number of synchrotron photons emitted) to make the timescale of this process comparable to other timescales in the system and observable within this test case. Further, to confirm this effect, the electron and positron populations were treated as independent within the \texttt{Diplodocus.jl} setup, with initially there being no positrons in the system.

The results of the time evolution of this setup are presented in \fref{fig: AM3 DIP ssc iso}. Just as was the case in \sref{subsec: AM3 sync} the two codes generate similar results. Both codes show a replenishing of the leptonic population at high momentum at late times due to pair-production from the high energy photon population, as evidenced by the appearance of a population of positrons. This replenished population forms the high energy tail of the leptonic populations which can also be observed to be cooling suggesting the rate of replenishment is lower than the rate of synchrotron and inverse-Compton cooling.

\begin{figure}[ht!]
    \centering
    \includegraphics[]{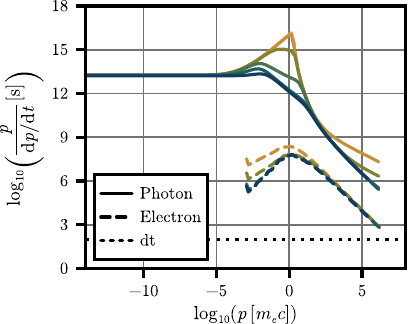}
    \caption{\label{fig: ssc timescale}Timescale for electron and photon momentum loss within \texttt{DiplodocusTransport.jl} for the synchrotron self-Compton test (\sref{subsec: AM3 SSC}). Line colours are matched to \fref{fig: AM3 DIP ssc iso} indicating simulation time and the horizontal line marked $dt$ indicates the size of the uniform time steps taken in this simulation.}
\end{figure}

However, it may be noted that once cooled to a momentum of around $p\approx 10^2\,[m_ec]$, via radiation reaction, the electron cooling rate begins to increase. as is demonstrated in \fref{fig: ssc timescale}, whereby the timescale of electron momentum loss begins to decrease with decreasing momenta, differing from the constant rate predicted from a radiation reaction force (see \fref{fig: rad react para perp timescale}). This is a feature resulting from large-angle scattering between high momentum photons and lower momentum electrons. In this regime, the electron's energy may change by a significant fraction with every photon collision, decreasing the cooling timescale. This explanation can be further justified by examining the photon loss rate also displayed in \fref{fig: ssc timescale}, where the loss rates at high energy approximately match\footnote{The photon loss timescale at high momentum also includes the effects of photon-photon annihilation to electron-positron pairs, but this is sub-dominant over the timescales examined in this test.} those for the low momentum electrons, indicating a common process (Compton scattering) linking the loss rates of these populations.

At late times ($t>\qty{e7}{\second}$), the population of high momentum photons may cause the rate of re-energisation of the leptonic population via Compton up-scattering to become dominant over radiation reaction cooling. However, computational intensity prevents examination of these late time effects (in this work) using the current version of \texttt{DiplodocusTransport.jl}. Unlike \texttt{AM\textsuperscript{3}}, which solves a linear transport equation (see \citet{KlingerEtAl_2024a}) \texttt{DiplodocusTransport.jl} includes full non-linear coupling between all species as described by \eref{eqn: flat transport w. coll}. This non-linearity stems from the binary collision matrix, which depends on the state of two incoming particles, allowing for the capture of more complicated interactions. However, a smaller time step is often required to accurately capture these interactions and maintain numerical stability. Thus requiring more time steps and increased computation time, compared to the optimised linear solvers used by \texttt{AM\textsuperscript{3}}. For example, \fref{fig: ssc timescale} shows that for accurate capture of the electron cooling over the entire duration of the simulation, the time step $dt$ was set to $\qty{e2}{\second}$, below the smallest loss timescale\footnote{The momentum loss timescale refers to the rate at which momentum is lost from each momentum bin, this does not directly imply cooling, as electrons can lose momentum from one bin by being up scattered to a higher momentum bin.} in the simulation at all times, requiring a total of $10^5$ time steps to reach the simulation end at $\qty{e7}{\second}$.

\subsubsection{\texorpdfstring{$\texttt{AM\textsuperscript{3}}$}{AM3} and \texttt{Diplodocus.jl} Runtime Comparison}\label{subsubsec: runtime comparison}

Table~\ref{tab:time} compares the runtime for the simulations conducted in Sections~\ref{subsec: AM3 sync} and \ref{subsec: AM3 SSC} with both \texttt{AM\textsuperscript{3}} and \texttt{Diplodocus.jl}. These simulations used different numbers of grid cells (\texttt{Diplodocus.jl} included $u$ grid cells that were not needed for these isotropic tests but are used in \sref{subsubsec: ani ssc}) and time stepping methods, hence their direct timings are not the best source of comparison. However, their runtime per time step per grid cell does provide a good source of comparison for how optimised these codes are. \texttt{Diplodocus.jl} appears to be approximately an order of magnitude slower when using this metric than \texttt{AM\textsuperscript{3}}, which is not surprising given the extensive optimisation that has been undertaken to reduce the computation required per time step in \texttt{AM\textsuperscript{3}} (see further, \cite{KlingerEtAl_2024a}).

\begin{table}[!h]
\centering
\caption{\label{tab:time}Simulation timing comparison between \texttt{AM\textsuperscript{3}} and \texttt{Diplodocus.jl} for Sections~\ref{subsec: AM3 sync} (Sync) and \ref{subsec: AM3 SSC} (SSC).}

\begin{tabular}{lcccc}
\toprule
\toprule
 & \multicolumn{2}{c}{\texttt{AM\textsuperscript{3}}} & \multicolumn{2}{c}{\texttt{Diplodocus.jl}} \\ 
 \midrule[0.5pt]
 & Sync & SSC & Sync & SSC \\ 
Time          & \qty{7.0}{s}  & \qty{48}{s} & \qty{3.6}{s} & \qty{23}{\hour} \\
Time steps    & 9999 & 1999  & 400 & $10^5$  \\
Grid cells                 & 790 & 1070 & 1512 & 2232 \\
Time/time step             & \qty{0.7}{ms} & \qty{0.024}{s}  & \qty{9.0}{ms} & \qty{0.84}{s} \\
Time/time step/grid cell   & \qty{0.88}{\micro s} & \qty{22}{\micro s} & \qty{6.0}{\micro s} & \qty{380}{\micro s}\\ 
\bottomrule
\end{tabular}
\end{table}

There are many improvements that can be made within the explicit time stepping scheme of \texttt{DiplodocusTransport.jl} described in \sref{subsec: scheme}. These steps only consists of dense matrix multiplication, therefore are very amenable to speed up by transitioning to GPU rather than CPU computation. In some cases these dense matrices are sparsely occupied and may be converted to sparse matrices for a reduction in computation and memory requirements. Further, as noted in \sref{subsec: AM3 SSC}, small time steps may be required to maintain stability, leading to long runtimes. An implicit time stepping scheme (see \sref{subsec: scheme}) may allow larger time steps to be taken while maintaining stability. These improvements are in development, and are planned to feature in \citetalias{PaperIII} and future versions of \texttt{DiplodocusTransport.jl}.

\begin{figure*}[!ht]
    \centering
    \vspace{5pt}
    \includegraphics[]{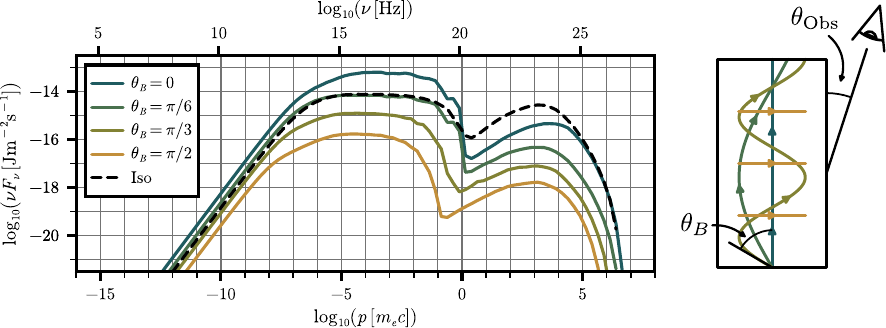}
    \caption{\label{fig: ssc B field angle dep}Spectral energy distribution, generated with \texttt{Diplodocus.jl}, as measured by an observer at an angle of $\theta_\text{Obs}=\ang{18}$ to the jet axis ($z$ cylindrical coordinate axis) at a simulation time of \qty{e5}{\second} using the setup described in \sref{subsubsec: ani ssc}. The magnetic field within the jet is taken to be either helical, with a constant pitch angle $\theta_B$, or isotropic.}
\end{figure*}

\newpage

\subsubsection{Anisotropic Synchrotron Self-Compton}\label{subsubsec: ani ssc}

The DIPLODOCUS framework was designed, in part, to extend isotropic, single-zone codes like \texttt{AM\textsuperscript{3}} to include anisotropic effects\,---\,thereby providing a framework through which the robustness of the assumption of isotropy within single-zone codes could be assessed. For example, within DIPLODOCUS, magnetic fields can have global structure rather than being simply being assumed to be isotropic on average. With advancements in polarimetric imaging of the jets emitted by astrophysical objects \citep{walkerEtAl2018,rodriguezEtAl2025,livingstonEtAl2025a} there is good evidence that these jets contain structured magnetic fields, likely helical in nature, lining up with what is suggested by the force-free configurations of black hole magnetospheres \citep{ChaelEtAl_2023,GellesEtAl_2025,TsunetoeEtAl_2025} and GRMHD simulations \citep{McKinneyGammie_2004,McKinney_2006,ChatterjeeEtAl_2019a}.

The effects of a helical field may be modelled by changing the spatial coordinates from spherical to cylindrical (\aref{app: cyl coords}) and rotating the local ortho-normal basis used for momentum-space coordinates, such that the local momentum $z$-axis is aligned with the magnetic field direction rather than the $z$-axis of the spatial cylindrical coordinate basis. For this test the magnetic field is taken to globally have a fixed pitch angle $\theta_B$ to the cylindrical $z$-axis and uniform magnitude $B=\qty{e-4}{\tesla}$ - corresponding to a generating current density $J^z=B\sin\theta_\text{B}/\mu_0r$. This field configuration and test setup\footnote{The spatial distribution of particles is still assumed to be homogenous in this test, therefore no particles are injected or escape except those described by the initial conditions.} is not necessarily realistic in the context of AGN jets, but provides an illustrative example of the effects that directed magnetic field may have on the observed spectral energy distribution of photons emanating from such a source.  

With a defined helical magnetic field structure, \fref{fig: ssc B field angle dep} shows the effect that is has on the observed spectral energy distribution compared with that of an isotropic field. The SEDs were measured at a fixed angle of $\theta_\text{Obs}=\qty{18}{\degree}$ to the cylindrical coordinate $z$-axis (\textit{jet} axis) at a simulation time of \qty{e5}{\second} (see \aref{app: obs flux} for details of how the SEDs were calculated to account for the cylindrical geometry). Five different magnetic field structures were examined, four helical with pitch angles $\theta_B=0,\pi/6,\pi/3,\pi/2$ and one isotropic, all with a field strength of $B=\qty{e-4}{\tesla}$. All configurations started with an identical power-law distribution of electrons and the same momentum-space grids used in \sref{subsec: AM3 sync}, except that the electron population was boosted by a modest bulk Lorentz factor of $\Gamma=2$. Note that this boosting is not along the global $z$-axis of the cylindrical coordinate system but along the local momentum $z$-axis, as such this bulk velocity is along the magnetic field direction---physically motivated by the fact that charged particles stream along field lines. Such boosting also ensures that the initial injected energy density (jet power) is identical between the different field configurations. The photon populations were then generated via synchrotron and synchrotron self-Compton.

It is evident from \fref{fig: ssc B field angle dep} that the observed spectral energy distributions for helical magnetic fields vary significantly from that of the isotropic case, by up to three orders of magnitude. The isotropic case consistently overestimates the observed inverse-Compton population of photons compared to that of the helical field cases, contrasted by the synchrotron bump which lies in-between the helical field cases. Within the helical field cases, there are differences between the overall spectral amplitude as well as the locations of features such as a spectral break in the synchrotron bump and peak of the inverse-Compton bump. There are several compounding effects generating these differences, such as: finite width of the momentum-space angular bins; Monte-Carlo integration noise; a line of sight effect as the magnetic field direction sweeps across the cylindrical surface of the emitting region (see \aref{app: obs flux}); beaming of the synchrotron emissions in the direction of electron propagation; and different electron cooling rates between the anisotropic and isotropic field configurations. What is clear however is that the inclusion of these anisotropic effects does make a difference on the observed spectral energy distribution. In depth study of these effects is beyond the scope of this paper, however, if studied further, these anisotropic effects may have a significant impact on inferred quantities, such as bulk Lorentz factor, jet power and observer angle, typically associated with modelling of blazars and other jetted astrophysical objects.

%% file: Sections/conclusion.tex
\section{Conclusion}\label{sec: conclusion}

In this paper, the numerical implementation of the DIPLODOCUS framework (as presented in \citetalias{EverettCotter_2025}) in the form of the code \texttt{Diplodocus.jl} has been described. This implementation consists of Monte-Carlo integration of anisotropic collision terms to generate collision matrices (\sref{sec:collisions}), including a novel approach to importance sampling of dependent variables, and the construction of flux matrices to describe the transport of particle distribution functions through phase space (\sref{sec:transport}). 

A series of test cases, focusing on micro-scale physics, i.e. forces and interactions that affect that transport particle distribution functions in momentum space but not physical space, are then presented in \sref{sec: test cases}. Validating the code in a range of scenarios both isotropic and anisotropic. In particular, comparisons are made to the single-zone emission modelling code \texttt{AM\textsuperscript{3}} \citep{KlingerEtAl_2024a}, demonstrating \texttt{Diplodocus.jl}'s ability to reproduce existing isotropic results (\ssref{subsec: AM3 sync} and \ref{subsec: AM3 SSC}) and expand upon them (even within a single-zone) to include angular dependent effects such as directed magnetic fields (\sref{subsubsec: ani ssc}), which can significantly effect observed spectra, likely impacting parametrisation of astrophysical sources (blazars in particular).

There remains a wide range of microphysical effects that could be implemented in \texttt{Diplodocus.jl}, e.g. magnetised turbulence, photon polarisation, synchrotron-self absorption and hadronic interactions all of which may be the focus of future work.

The next paper in the series, \citetalias{PaperIII}, will focus on the testing of macro-scale physics. This includes the spatial advection of particle distributions though a range of spacetime configurations with and without the inclusion of micro-scale physics, demonstrating the full mesoscopic range of \texttt{Diplodocus.jl} and the codes performance when scaled to larger compute systems.

%% file: Appendicies/differential_cross_sections.tex
\section{Differential Cross Sections}\label{app: diff x sections} 

Here are presented the differential cross sections used in evaluating collision terms. For binary interactions, $\mathfrak{ab}\leftrightharpoons \mathfrak{cd}$, cross sections are written in a Lorentz invariant form using the Mandelstam variables $S,T,\text{and }U$ defined by 
\begin{equation}\label{eqn: mandelstram variables}
    S = \left(\bm{p}_\mathfrak{a}+\bm{p}_\mathfrak{b}\right)^2=\left(\bm{p}_\mathfrak{c}+\bm{p}_\mathfrak{d}\right)^2,
    \quad
    T = \left(\bm{p}_\mathfrak{a}-\bm{p}_\mathfrak{c}\right)^2=\left(\bm{p}_\mathfrak{d}-\bm{p}_\mathfrak{b}\right)^2,
    \quad
    U = \left(\bm{p}_\mathfrak{a}-\bm{p}_\mathfrak{d}\right)^2=\left(\bm{p}_\mathfrak{c}-\bm{p}_\mathfrak{b}\right)^2.
\end{equation}

\subsection{Hard Sphere Collisions \texorpdfstring{$\circ\circ\leftrightharpoons\circ\circ$}{○○⇌○○}}\label{app: hard sphere x-sec}
For the collision of idealised spheres of radius $R_\circ$ and mass $m_\circ$, the differential and total cross sections are given by:
\begin{subequations}
\begin{align}
    \label{eqn: diff x section hard sphere}
    \frac{\mathrm{d}\sigma_{\circ\circ\rightarrow\circ\circ}}{\mathrm{d}T}&=\frac{\pi R_\circ^2}{S-4m_\circ^2},
    \\
    \label{eqn: total x section hard sphere}
    \sigma_{\circ\circ\rightarrow\circ\circ}&=\frac{\pi R_\circ^2}{2},
\end{align}
\end{subequations}
where an additional factor of $1/2$ has been included in the total cross section to account for the indistinguishability of output states. 

\subsection{Electron(Positron)-Photon Scattering \texorpdfstring{$e\gamma\leftrightharpoons e\gamma$}{eγ⇌eγ}}\label{app: compt x-sec}
For the Compton scattering of an electron (positron) and a photon, the differential and total cross sections are given by \citep[][Eqs.~(86.6) and (86.16)]{BerestetskiiEtAl_1982}: 
\begin{subequations}
\begin{align}
    \frac{\mathrm{d}\sigma_{e\gamma\rightarrow e\gamma}}{\mathrm{d}T} &= \frac{3\sigma_\text{T}m_{e}^2}{(S-m_{e}^2)^2}\left[\left(\frac{m_{e}^2}{S-m_{e}^2}+\frac{m_{e}^2}{U-m_{e}^2}\right)^2+\left(\frac{m_{e}^2}{S-m_{e}^2}+\frac{m_{e}^2}{U-m_{e}^2}\right)-\frac{1}{4}\left(\frac{S-m_{e}^2}{U-m_{e}^2}+\frac{U-m_{e}^2}{S-m_{e}^2}\right)\right]
    \\
    \sigma_{e\gamma\rightarrow e\gamma}(s) &= \frac{3\sigma_\text{T}m_{e}^2}{4(S-m_{e}^2)}\left[\left(1-\frac{4m_{e}^2}{\left(S-m_{e}^2\right)}-\frac{8m_e^4}{\left(S-m_{e}^2\right)^2}\right)\log\left(\frac{S}{m_{e}^2}\right)+\frac{1}{2}+\frac{8m_{e}^2}{S-m_{e}^2}-\frac{m_{e}^4}{2S^2}\right]
\end{align}
\end{subequations}

\subsection{Electron-Positron Annihilation to Two Photons \texorpdfstring{$e^{-}e^{+} \rightarrow \gamma\gamma$}{ee→γγ}}\label{app: pair anni x-sec}
For the annihilation of an electron-positron pair to two photons, the differential and total cross sections are given by \citep[][Eqs.~(88.4) and (88.6)]{BerestetskiiEtAl_1982}:
\begin{subequations}
\begin{align}
    \frac{\mathrm{d}\sigma_{e^+e^-\rightarrow\gamma\gamma}}{\mathrm{d}T} &= -\frac{3\sigma_\text{T}m_{e}^2}{S(S-4m_{e}^2)}\left[\left(\frac{m_{e}^2}{T-m_{e}^2}+\frac{m_{e}^2}{U-m_{e}^2}\right)^2+\left(\frac{m_{e}^2}{T-m_{e}^2}+\frac{m_{e}^2}{U-m_{e}^2}\right)-\frac{1}{4}\left(\frac{T-m_{e}^2}{U-m_{e}^2}+\frac{U-m_{e}^2}{T-m_{e}^2}\right)\right]
    \\
    \sigma_{e^+e^-\rightarrow\gamma\gamma} &= \frac{3\sigma_\text{T}m_{e}^2}{4S^2(S-4m_{e}^2)}\left[(S^2+4Sm_{e}^2-8m_{e}^4)\log\left(\frac{\sqrt{S}+\sqrt{S-4m_{e}^2}}{\sqrt{S}-\sqrt{S-4m_{e}^2}}\right)-(S+4m_{e}^2)\sqrt{S(S-4m_{e}^2)}\right]
\end{align}
\end{subequations}

\subsection{Photon-Photon Annihilation to Electron-Positron Pair \texorpdfstring{$\gamma\gamma\rightarrow e^{-}e^{+}$}{γγ→ee}}\label{app: photon anni x-sec}
For the annihilation of two photons to an electron-positron pair, the differential and total cross sections are given by $\frac{\mathrm{d}\sigma_{\gamma\gamma\rightarrow e^+e^-}}{\mathrm{d}T}=\frac{S-4m_e^2}{S}\frac{\mathrm{d}\sigma_{e^+e^-\rightarrow\gamma\gamma}}{\mathrm{d}T}$:
\begin{subequations}
\begin{align}
    \frac{\mathrm{d}\sigma_{\gamma\gamma\rightarrow e^+e^-}}{\mathrm{d}T} &= -\frac{3\sigma_\text{T}m_{e}^2}{S^2}\left[\left(\frac{m_{e}^2}{T-m_{e}^2}+\frac{m_{e}^2}{U-m_{e}^2}\right)^2+\left(\frac{m_{e}^2}{T-m_{e}^2}+\frac{m_{e}^2}{U-m_{e}^2}\right)-\frac{1}{4}\left(\frac{T-m_{e}^2}{U-m_{e}^2}+\frac{U-m_{e}^2}{T-m_{e}^2}\right)\right]
    \\
    \sigma_{\gamma\gamma\rightarrow e^+e^-} &= \frac{3\sigma_\text{T}m_{e}^2}{2S^3}\left[(S^2+4Sm_{e}^2-8m_{e}^4)\log\left(\frac{\sqrt{S}+\sqrt{S-4m_{e}^2}}{\sqrt{S}-\sqrt{S-4m_{e}^2}}\right)-(S+4m_{e}^2)\sqrt{S(S-4m_{e}^2)}\right]
\end{align}
\end{subequations}

%% file: Appendicies/radreactandsync.tex
\section{Radiation Reaction and Synchrotron Emissions}\label{app: rad react sync}

\subsection{Radiation Reaction}\label{app: rad react}

The reaction force on a charged particle undergoing a time varying acceleration due to an external force is given by the Abraham-Lorentz-Dirac force \citep{Abraham_1905,Lorentz_1892,Dirac_1938}. In a local orthonormal basis $\bm{e}_a$, this force is given by 
\begin{equation}
    F^a=\frac{\mu_0q^2}{6\pi mc}\left(\eta^{ab}+\frac{p^a p^b}{m^2c^2}\right)\frac{\mathrm{d}^2}{\mathrm{d}^2\tau}p_b.
\end{equation}
This depends on the second derivative of the particle's momenta making the system difficult to evolve, however assuming the radiation reaction force is small compared to some external force $\bm{F}_\text{ext}$ a reduction of order (Landau-Lifshitz approximation) \citep{PoissonEtAl_2011,LandauLifshits_1975}, can take place with $\mathrm{d}p^a/\mathrm{d}\tau\approx\bm{F}_\text{ext}$. Considering an external force from a magnetic field $F^{a}_\text{ext}=qF^{ab}p_b/m$, the Lorentz force, with $F^{ab}$ being the electromagnetic field tensor. If there is no electric field and the magnetic field is aligned to the local $\bm{e}_3$ direction, the field tensor is given by $F^{ab}=\epsilon^{ab03}B$, where $\bm{\epsilon}$ is the completely antisymmetric Levi-Civita tensor. Under the effect of this external force, the components of the radiation reaction force are given by:
\begin{subequations} \label{eqn: rad react force}
    \begin{align}
        F^0&=-\frac{(q/q_e)^4B^2\sigma_\text{T}}{\mu_0(m/m_e)^3m_ec}\left(\frac{(p)^2\sin^2\theta}{m^2c^2}\right)p^0,
        \\
        F^1&=\frac{(q/q_e)^4B^2\sigma_\text{T}}{\mu_0(m/m_e)^3m_ec}\left(-1-\frac{(p)^2\sin^2\theta}{m^2c^2}\right)p^1,
        \\
        F^2&=\frac{(q/q_e)^4B^2\sigma_\text{T}}{\mu_0(m/m_e)^3m_ec}\left(-1-\frac{(p)^2\sin^2\theta}{m^2c^2}\right)p^2,
        \\
        F^3&=\frac{(q/q_e)^4B^2\sigma_\text{T}}{\mu_0(m/m_e)^3m_ec}\left(-\frac{(p)^2\sin^2\theta}{m^2c^2}\right)p^3.
    \end{align}
\end{subequations}
The $0$ component can be related to the classical synchrotron power (i.e. rate of energy loss as measured by a stationary observer), given by
\begin{equation}\label{eqn: rad react power}
    P=\frac{mc^2 F^0}{p^0}=-\frac{(q/q_e)^4B^2\sigma_\text{T}c}{\mu_0(m/m_e)^2}\left(\frac{(p)^2\sin^2\theta}{m^2c^2}\right),
\end{equation}

\subsection{Synchrotron}\label{app: synchrotron}
A charged particle (denoted by $\mathfrak{a}$), of charge $q_\mathfrak{a}$ and mass $m_\mathfrak{a}$, orbiting a magnetic field line of strength $B$ in a helical path with pitch angle $\theta_\mathfrak{a}$ and momentum $p_\mathfrak{a}$, will experience the radiation reaction force as described by \eref{eqn: rad react force} and emit synchrotron\footnote{No constraints on the charged particles momentum are made in this work, hence we use ``synchrotron'' as a catch all term for both non-relativistic (cyclotron) and relativistic emissions.} photons (denoted by $\mathfrak{c})$. The observed angular power spectrum of these photons is given by \citep{Schott_1912,SokolovTernov_1968,SokolovEtAl_1969,LandauLifshits_1975}:
\begin{equation}\label{eqn: sync angular power spectrum}
    \frac{\mathrm{d}P_n}{\mathrm{d}\Omega_\mathfrak{c}}=\frac{q_\mathfrak{a}^2}{q_e^4}\frac{3m_e^2c^3\sigma_\text{T}}{4\pi\mu_0\hbar^2}\frac{(p_\mathfrak{c})^2}{p^0_\mathfrak{a}\left(p^0_\mathfrak{a}-p_\mathfrak{a}\cos\theta_\mathfrak{a}\cos\theta_\mathfrak{c}\right)}\left[\left(\frac{p^0_\mathfrak{a}\cos\theta_\mathfrak{c}-p_\mathfrak{a}\cos\theta_\mathfrak{a}}{\sin\theta_\mathfrak{c}}\right)^2J_n(x)^2+p_\mathfrak{a}^2\sin^2\theta_\mathfrak{a} J'_{n}(x)^2\right],
\end{equation}
where $n$ is the fundamental harmonic of the emission $n=\omega_\mathfrak{c}/\omega_*=p_\mathfrak{c}/p_*$, with the fundamental ``energy'' $p_*$ given by 
\begin{equation}
    p_*=\frac{\hbar q_\mathfrak{a}B}{\left(p^0_\mathfrak{a}-p_\mathfrak{a}\cos\theta_\mathfrak{a}\cos\theta_\mathfrak{c}\right)},
\end{equation}
$J_n(x)$ are the Bessel functions of the first kind, with argument $x=np_\mathfrak{a}\sin\theta_\mathfrak{a}\sin\theta_{\mathfrak{c}}/\left(p^0_\mathfrak{a}-p_\mathfrak{a}\cos\theta_\mathfrak{a}\cos\theta_\mathfrak{c}\right)$ and $J'_n(x)=\partial{ J_n(x)}/\partial x=\left(J_{n-1}(x)-J_{n+1}(x)\right)/2$.
By summing over the fundamental harmonics and integrating over the solid angle \citep{SokolovEtAl_1969}, the total power emitted is given by 
\begin{equation}\label{eqn: sync emission power}
    P = \frac{(q_\mathfrak{a}/q_e)^4B^2\sigma_\text{T}c}{\mu_0(m_\mathfrak{a}/m_e)^2}\left(\frac{(p_\mathfrak{a})^2\sin^2\theta_\mathfrak{a}}{m_\mathfrak{a}^2c^2}\right),
\end{equation}
which is identical to \eref{eqn: rad react power}, hence the rate at which the emitting particle is loosing energy is equivalent to the total radiated energy of photons.

The total number of photons observed to be emitted per unit time is given by summing the angular power spectrum \eref{eqn: sync angular power spectrum} over all harmonics and divided by the energy of the emitted photons
\begin{equation}
    \frac{\mathrm{d}N_{\mathfrak{a}\rightarrow\mathfrak{a}\mathfrak{c}}}{\mathrm{d}t\mathrm{d}\Omega_\mathfrak{c}}=\sum_n\frac{1}{p_\mathfrak{c}c}\frac{\mathrm{d}P_n}{\mathrm{d}\Omega_\mathfrak{c}}.
\end{equation}
Using that $n=\frac{p_\mathfrak{c}}{p_*}$ this can be re-written as an integral over $p_\mathfrak{c}$
\begin{equation}
    \frac{\mathrm{d}N_{\mathfrak{a}\rightarrow\mathfrak{a}\mathfrak{c}}}{\mathrm{d}t\mathrm{d}\Omega_\mathfrak{c}}=\int\sum_n\frac{1}{p_\mathfrak{c}c}\frac{\mathrm{d}P_n}{\mathrm{d}\Omega_\mathfrak{c}}\delta_{n,p_\mathfrak{c}/p_*}\frac{\mathrm{d}p_\mathfrak{c}}{p_*},
\end{equation}
where $\delta$ here is the Kronecker delta. 

For evaluation of synchrotron emissions via \texttt{DiplodocusCollisions.jl}, the gain term (\eref{eqn: emissive gain term}) for this emissive interaction is then given by
\begin{equation}\label{eqn: gain term synchrotron}
    G_{\mathfrak{a}\rightarrow\mathfrak{a}\mathfrak{c}} = \sum_n \frac{q_\mathfrak{a}}{B}\frac{3m_e^2c^3\sigma_\text{T}}{4\pi\mu_0\hbar^3q_e^4}\frac{p_\mathfrak{c}}{p^0_\mathfrak{a}}\left[\left(\frac{p^0_\mathfrak{a}\cos\theta_\mathfrak{c}-p_\mathfrak{a}\cos\theta_\mathfrak{a}}{\sin\theta_\mathfrak{c}}\right)^2J_n(x)^2+p_\mathfrak{a}^2\sin^2\theta_\mathfrak{a} J'_{n}(x)^2\right]\delta_{n,p_\mathfrak{c}/p_*}
\end{equation}
It is standard to assume that for $n\gg1$ the emission spectrum is taken as continuous, allowing the removal of the sum and Kronecker delta. This is well justified for relativistic $p_\mathfrak{a}\gg m_\mathfrak{a}c$ where the spectrum is dominated by higher order harmonics \citep{Pacholczyk_1970,RybickiLightman_2004} but as this work makes no specification as to the range of $p_\mathfrak{a}$ considerations must be made. These considerations are achieved by choosing a tolerance for which $p_\mathfrak{c}/p_*\approx \mathbb{Z}$. The error between $p_\mathfrak{c}/p_*$ and the nearest integer can be calculated using 
\begin{equation}
    err = \frac{p_\mathfrak{c}/p_*-\left\lfloor p_\mathfrak{c}/p_* \right\rceil}{\left\lfloor p_\mathfrak{c}/p_* \right\rceil},
\end{equation}
where $\left\lfloor p_\mathfrak{c}/p_* \right\rceil$ is the value of $p_\mathfrak{c}/p_*$ rounded to the nearest integer. If $err\leq5\times10^{-m}$, the maximum difference between an un-rounded value and its closest integer is $0.5$, hence this criterion is always true for $n\geq10^{m-1}$. This error criterion is used for the tolerance on $\delta_{n,p_\mathfrak{c}/p_*}$, if $err\leq5\times10^{-3}$ the Kronecker delta is taken to have the value of unity, and the sum in \eref{eqn: gain term synchrotron} only has a single value, other wise the Kronecker delta is equal to zero and so is the gain term. 

For numerical stability, if $n>10^6$ and $y=x/n>1-10^{-3}$ i.e. the argument of the Bessel functions is large, the following approximations are made: 
\begin{subequations}
    \begin{align}
    J_{n}(x) &\approx \frac{\sqrt{1-y^2}}{\sqrt{3}\pi}\left[K_{1/3}\left(z\right)+\frac{1-y^2}{10}\left(K_{1/3}(z)-2n\left(1-y^2\right)^{3/2}K_{2/3}(z)\right)\right],
    \\
    J'_{n}(x) &\approx \frac{1-y^2}{\sqrt{3}\pi}\left[K_{2/3}\left(z\right)+\frac{1-y^2}{5}\left(2K_{2/3}(z)-\left(\frac{1}{n\left(1-y^2\right)^{3/2}}+n\left(1-y^2\right)^{3/2}\right)K_{1/3}(z)\right)\right],
    \end{align}
\end{subequations} 
where $z=n\left(1-y^2\right)^{3/2}/3$, and though being approximations, still maintain the total synchrotron power given by \eref{eqn: sync emission power} \citep{SokolovEtAl_1969}.

%% file: Appendicies/scattering_limits.tex
\section{Binary Scattering Angle Limits}\label{app: angle limits}
For binary interactions, there may be a maximum physically allowable scattering angle $\theta_\text{max}$ in the observer frame with respect to the centre of momentum velocity direction. 

By Lorentz transformations, the momenta in the centre of momentum and observer frames (primed and un-primed variables respectively) are related by
\begin{equation}
p\cos\theta = p^*\cos\theta^*\cosh \rho^*+E^*\sinh \rho^*, \quad
p\sin\theta = p^*\sin\theta^*,
\end{equation}
where $p^*=\mathcal{F}_\mathfrak{ab}/\sqrt{S}$ and $E^*=\sqrt{(p^*)^2+m^2}$ are the centre of momentum frame momentum and energy of the particle, and $\rho^*$ is the rapidity of that frame with respect to the observer frame. The functions $\mathcal{F}_{\mathfrak{a}\mathfrak{b}}=\sqrt{\left(S-(m_\mathfrak{a}+m_\mathfrak{b})^2\right)\left(S-(m_\mathfrak{a}-m_\mathfrak{b})^2\right)}/2$, $\rho^*=\cosh^{-1}\left(\left(p^0_\mathfrak{a}+p^0_\mathfrak{b}\right)/\sqrt{S}\right)$, $S$ (\eref{eqn: mandelstram variables}), and $p^*$ are only depend on the incoming state variables, as measured in the observer frame. 

By eliminating the observer frame momentum $p$ and maximising $\theta$ with respect to $\theta^*$, the maximum observer frame scattering angle is obtained: 
\begin{equation}
    \tan\theta_\text{max}=\frac{p^*}{\sqrt{m^2\sinh^2 w^*-(p^*)^2}}, \quad \cos\theta_\text{max}=\sqrt{1-\left(\frac{p^*}{m\sinh w^*}\right)^2}.
\end{equation}

For good sampling of angles in the observer frame, the effective rapidity $\tilde{\rho}$ is defined as that which maps $\tilde{\theta}=\pi/2\rightarrow \theta_\text{max}$, i.e. if $\tilde{\theta}$ are sampled uniformly, half of the samples will lie with $\theta<\theta_\text{max}$ and therefore that region will be well sampled. The Doppler boosting formula for angle cosines is given by 
\begin{equation}
    \cos\theta=\frac{\cos\tilde\theta+\tanh\tilde \rho}{1+\cos\tilde\theta\tanh\tilde \rho},
\end{equation}
with $\tilde\theta=\pi/2$ and $\theta=\theta_\text{max}$ the value of $\tilde \rho$ is given by 
\begin{equation}\label{eqn: tilde rap}
    \tilde \rho=\tanh^{-1}\left(\sqrt{1-\left(\frac{p^*}{m\sinh w^*}\right)^2}\right).
\end{equation}
 

%% file: Appendicies/minkowski_coordinates.tex
\section{Coordinate Definitions}\label{app: coords}
This section follows the nomenclature of \citetalias{EverettCotter_2025} Appendix B, for the definition of the coordinates, tetrad and Ricci rotation coefficients.

\subsection{Cartesian Minkowski Coordinates}\label{app: cart coords}
For cylindrical Minkwoski coordinates $\{t,x,y,z\} = \{t,x,y,z\}$ and the metric is given by: 
\begin{equation}\label{eqn: cart metric}
\bm{g} = -\bm{\mathrm{d}}t\otimes\bm{\mathrm{d}}t + \bm{\mathrm{d}}x\otimes\bm{\mathrm{d}}x+\bm{\mathrm{d}}y\otimes\bm{\mathrm{d}}y+\bm{\mathrm{d}}z\otimes\bm{\mathrm{d}}z.
\end{equation}

In this coordinate system, a stationary observer has components $n_a=(-1,0,0,0)$ and the local orthonormal bases are taken to be aligned with the coordinate axis such that the tetrad described by Eq.~(44) of \citetalias{EverettCotter_2025} is:
\begin{equation}\label{eqn: cart tetrad}
    e_a^{~\alpha}=
    \begin{pmatrix}
    -1 & 0 & 0 & 0 
    \\
    0 & 1 & 0 & 0 
    \\
    0 & 0 & 1 & 0
    \\
    0 & 0 & 0 & 1
    \end{pmatrix},
\end{equation}
The Ricci rotation coefficients for this coordinate system and tetrad are all equal to zero $\Gamma^{a}_{~bc}=0$.

\subsection{Cylindrical Minkowski Coordinates}\label{app: cyl coords}
For cylindrical Minkwoski coordinates $\{t,x,y,z\} = \{t,\rho,\vartheta,z\}$ and the metric is given by: 
\begin{equation}\label{eqn: cyl metric}
\bm{g} = -\bm{\mathrm{d}}t\otimes\bm{\mathrm{d}}t + \bm{\mathrm{d}}\rho\otimes\bm{\mathrm{d}}\rho+\rho^2\bm{\mathrm{d}}\vartheta\otimes\bm{\mathrm{d}}\vartheta+\bm{\mathrm{d}}z\otimes\bm{\mathrm{d}}z.
\end{equation}

In this coordinate system, a stationary observer has components $n_a=(-1,0,0,0)$ and the standard tetrad described by Eq.~(44) of \citetalias{EverettCotter_2025} is:
\begin{equation}\label{eqn: cyl tetrad}
    e_a^{~\alpha}=
    \begin{pmatrix}
    -1 & 0 & 0 & 0 
    \\
    0 & \cos\alpha\cos\gamma-\sin\alpha\cos\beta\sin\gamma & \frac{-\sin\alpha\cos\gamma-\cos\alpha\cos\beta\sin\gamma}{\rho} & \sin\beta\sin\gamma 
    \\
    0 & \sin\alpha\cos\beta\cos\gamma+\cos\alpha\sin\gamma & \frac{\cos\alpha\cos\beta\cos\gamma-\sin\alpha\sin\gamma}{\rho} & -\sin\beta\cos\gamma
    \\
    0 & \sin\alpha\sin\beta & \frac{\cos\alpha\sin\beta}{\rho} & \cos\beta
    \end{pmatrix},
\end{equation}
where $\alpha,\beta$ and $\gamma$ are a set of three Euler angles that describe a ``z-x-z'' rotation of the spatial orthonormal basis vectors. For the case where $\alpha=\gamma=0$ and $\beta=\text{constant}$, i.e. the stationary observers are aligned with a helical field with constant pitch angle $\beta$. The non-zero Ricci rotation coefficients for this coordinate system and tetrad are given by:
\begin{equation}
\begin{split}
\Gamma^{1}_{~22} &= -\frac{\cos^2\beta}{\rho}, \quad \Gamma^{1}_{~23} = \Gamma^{1}_{~32} = -\frac{\sin\beta\cos\beta}{\rho}, \quad \Gamma^{1}_{~33} = -\frac{\sin^2\beta}{\rho}, \quad \Gamma^{2}_{~12} = \frac{\cos^2\beta}{\rho},  
\\
\Gamma^{2}_{~13} &= \frac{\sin\beta\cos\beta}{\rho}, \quad\Gamma^{3}_{~12} = \frac{\sin\beta\cos\beta}{\rho}, \quad \Gamma^{3}_{~13} = \frac{\sin^2\beta}{\rho}.
\end{split}
\end{equation}

\subsection{Spherical Minkowski Coordinates}\label{app: sph coords}
For cylindrical Minkwoski coordinates $\{t,x,y,z\} = \{t,r,\theta,\psi\}$ and the metric is given by: 
\begin{equation}\label{eqn: sph metric}
\bm{g} = -\bm{\mathrm{d}}t\otimes\bm{\mathrm{d}}t + \bm{\mathrm{d}}r\otimes\bm{\mathrm{d}}r+r^2\bm{\mathrm{d}}\theta\otimes\bm{\mathrm{d}}\theta+r^2\sin^2\theta\bm{\mathrm{d}}\psi\otimes\bm{\mathrm{d}}\psi.
\end{equation}

In this coordinate system, a stationary observer has components $n_a=(-1,0,0,0)$ and the local orthonormal bases are taken to be aligned with the coordinate axis such that the tetrad described by Eq.~(44) of \citetalias{EverettCotter_2025} is:
\begin{equation}\label{eqn: sph tetrad}
    e_a^{~\alpha}=
    \begin{pmatrix}
    -1 & 0 & 0 & 0 
    \\
    0 & 1 & 0 & 0 
    \\
    0 & 0 & r^2 & 0
    \\
    0 & 0 & 0 & r^2\sin^2\theta
    \end{pmatrix},
\end{equation}
The non-zero Ricci rotation coefficients for this coordinate system and tetrad are given by:
\begin{equation}
\Gamma^{1}_{~22} = \Gamma^{1}_{~33} = -\frac{1}{r}, \quad \Gamma^{2}_{~12} = \Gamma^{3}_{~13} = \frac{1}{r} \quad \Gamma^{2}_{~33} = -\frac{\cot\theta}{r}, \quad \Gamma^{3}_{~23} = \frac{\cot\theta}{r}.
\end{equation}

%% file: Appendicies/test_cases_num_eng.tex
\section{Test Cases: Number and Energy Density Conservation}\label{app: num eng den}

This section displays particle number and energy density as a function of time for the test cases presented in \sref{sec: test cases}.

\subsection{Elastically Colliding Spheres}\label{app: hard sphere num eng den}

\begin{figure}[!ht]
    \centering
    \includegraphics[scale=0.9]{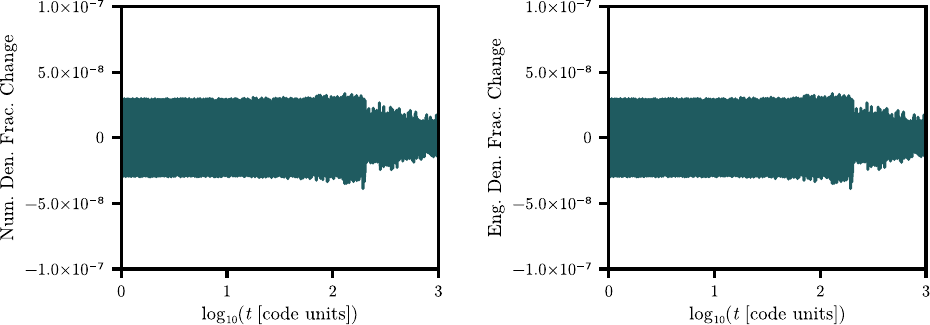}
    \caption{Fractional change in sphere number (left) and energy (right) density per 100 time steps a function of time for the evolution of hard spheres under elastic collisions as described in \sref{subsec: hard spheres}.}
    \label{fig: hard sphere frac num den frac eng den}
\end{figure}

\fref{fig: hard sphere frac num den frac eng den} shows the change in hard sphere number and energy density as a function of time for the test described in \sref{subsec: hard spheres}. The two plots in \fref{fig: hard sphere frac num den frac eng den} are almost identical as energy density $e$ is proportional to number density $n$, with them being related by $e=\bar{p^0}n$, where $\bar{p^0}$ is the mean energy per particle.

\subsection{Electron Gyration and Drift}\label{app: ExB num eng den}

\begin{figure}[!ht]
    \centering
    \includegraphics[scale=0.9]{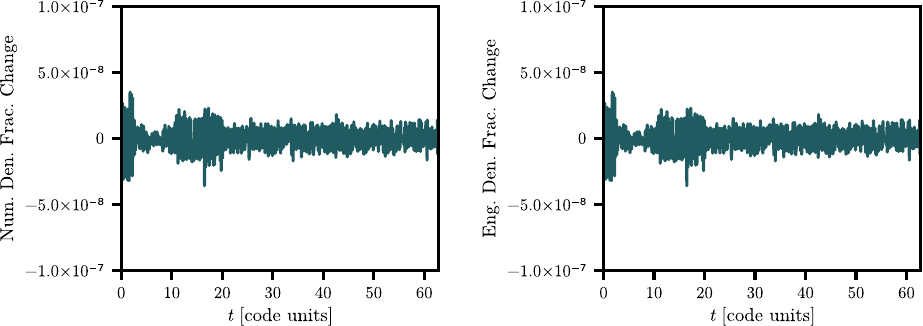}
    \caption{Fractional change in electron number density (left) and energy density (right) per time steps as a function of time over 10 gyrations of an distribution of electrons experiencing a Lorentz force from a uniform magnetic field, as described in \sref{subsubsec: ExB gyration}.}
    \label{fig: ExB no E frac num den eng den}
\end{figure}

\begin{figure}[!ht]
    \centering
    \includegraphics[scale=0.9]{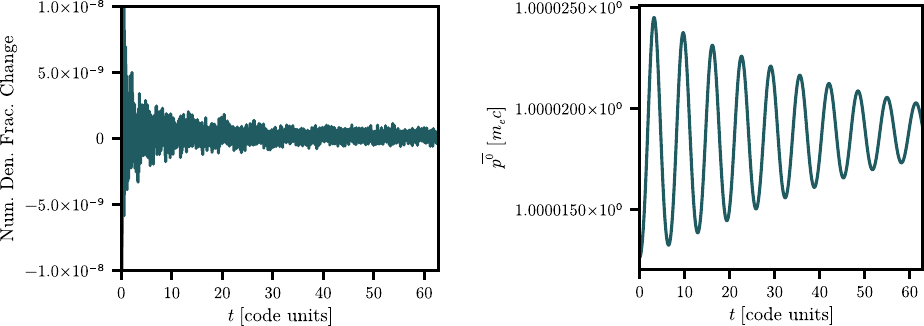}
    \caption{Fractional change in electron number density per time step (left) and mean energy per electron (right) as a function of time over 10 gyrations of a distribution of electrons experiencing a Lorentz force from a uniform magnetic field and uniform, perpendicular electric field, as described in \sref{subsubsec: ExB drift}.}
    \label{fig: ExB frac num den eng den}
\end{figure}

\fsref{fig: ExB no E frac num den eng den} and \ref{fig: ExB frac num den eng den} show the change in electron number and energy density as a function of time for the tests described in \ssref{subsubsec: ExB gyration} and \ref{subsubsec: ExB drift} respectively. As there are no collision or emission terms, number density is conserved to machine precision in both cases. This is also true for energy density in the case of electron gyration presented in \sref{subsubsec: ExB gyration} and \fref{fig: ExB no E frac num den eng den}. Whereas, in the case of electron drift presented in \sref{subsubsec: ExB drift} and \fref{fig: ExB frac num den eng den} the mean energy per particle follows the expected sinusoidal pattern centred on the mean energy per gyration as electrons are accelerated by the electric field for half a gyration and then decelerated for the second half. The amplitude of these oscillations in mean energy decrease in time as the population spreads out over the $\phi$ coordinate (see \fref{fig: ExB phi dis}). However, at late times ($>10$ gyrations for this test) the numerical diffusion in the $p$ coordinate reaches bounds of the momentum domain, preventing accurate gyration and drift.

\subsection{Radiation Reaction}\label{app: rad react num eng den}

\begin{figure}[!ht]
    \centering
    \includegraphics[scale=0.9]{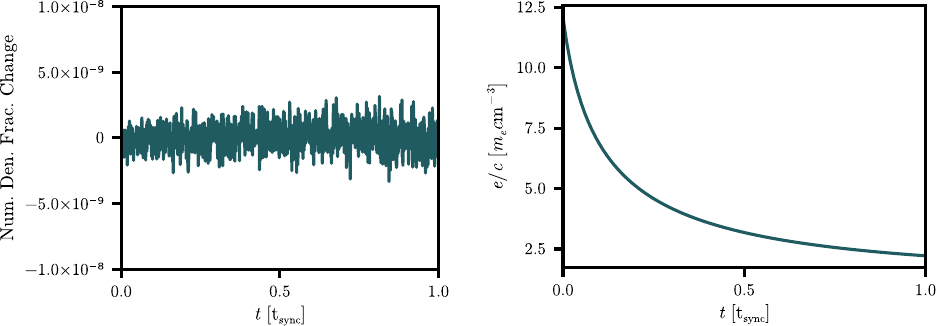}
    \caption{Fractional change in electron number density per 10 time steps (left) and energy density (right) as a function of time for the evolution of electrons experiencing a radiation reaction force as described in \sref{subsec: ele rad raction}.}
    \label{fig: rad react frac num den eng den}
\end{figure}

\fref{fig: rad react frac num den eng den} show the change in number and energy density of electrons as a function of time for the test described in \sref{subsec: ele rad raction}. Energy density is decreasing with time

\subsection{Synchrotron and Synchrotron Self-Compton}\label{app: sync and SSC num eng}

\begin{figure}[!ht]
    \centering
    \includegraphics[scale=0.9]{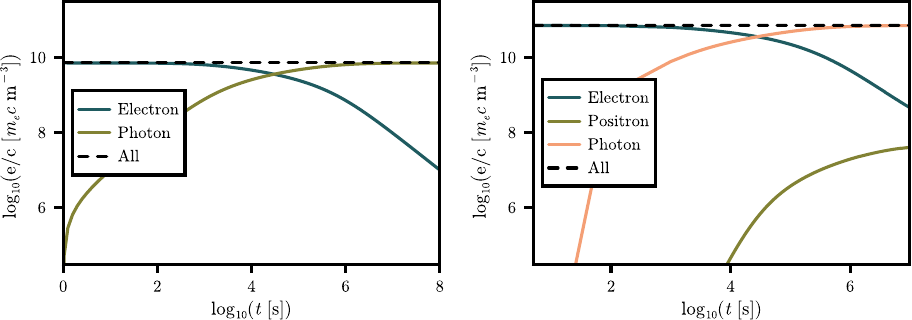}
    \caption{Energy density for electrons and photons as a function of time for the synchrotron test case presented in \sref{subsec: AM3 sync} (left) and for the synchrotron self-Compton test presented in \sref{subsec: AM3 SSC} (right, additionally including the positron energy density).}
    \label{fig: sync eng dens}
\end{figure}

\fref{fig: sync eng dens} show the change in energy density as a function of time for the tests described in \ssref{subsec: AM3 sync} and \ref{subsec: AM3 SSC}. Energy conservation is guaranteed between radiation reaction and synchrotron emissions by a corrective term (see \aref{app: sync num eng}). The binary collision terms that generate the (inverse-)Compton scattering and the production of electron-positron pairs by photon annihilation also conserve number and energy density to machine precision by applying a corrective scaling to the gain array elements to counter noise introduced by inaccurate Monte-Carlo sampling (see \sref{subsubsec: binary correction}).

\begin{figure}[!ht]
    \centering
    \includegraphics[scale=0.9]{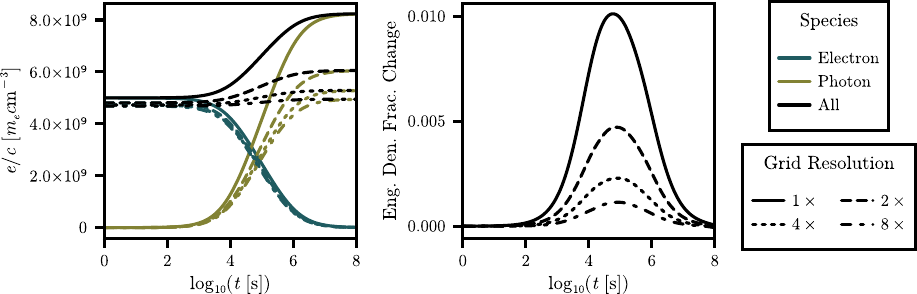}
    \caption{\label{fig: sync eng dens wo correction}Energy density for the synchrotron test case presented in \sref{subsec: AM3 sync}, \textbf{without energy correction}, as a function of time (left), and as a function of momentum grid resolution, where the solid line has a resolution $1\times$ that of \sref{subsec: AM3 sync}. Fractional change in total energy density is also shown (right).}
\end{figure}

\subsection{Synchrotron Energy Conservation Without Correction}\label{app: sync num eng}

As mentioned in \sref{subsec: AM3 sync}, without a corrective term on the synchrotron emission matrix, the synchrotron test displayed in \fref{fig: AM3 DIP ssc iso} would not conserve energy between electrons and photons. Electrons are cooled via a radiation reaction force, implemented via a force flux term in \texttt{Diplodocus.jl}, whereas synchrotron photons are implemented via a Monte-Carlo sampled emission matrix. As these two methods conserve energy theoretically (as described in \aref{app: rad react sync}) but the finite grid resolutions and noise introduced by the Monte-Carlo integration may cause deviations from energy conservation. 

\fref{fig: sync eng dens} demonstrates this change in energy density over time, without a corrected term applied, and further examines the same test taken at increasing momentum grid resolutions. It is evident that energy is better conserved for higher grid resolutions, with the maximum fractional changed decreasing linearly with grid resolution, indicating perfect conservation in the limit of an infinitely resolved grid.

%% file: Appendicies/obs_flux.tex
\section{Observed Spectrum}\label{app: obs flux}

\begin{figure}[!ht]
    \centering
    \includegraphics[scale=0.9]{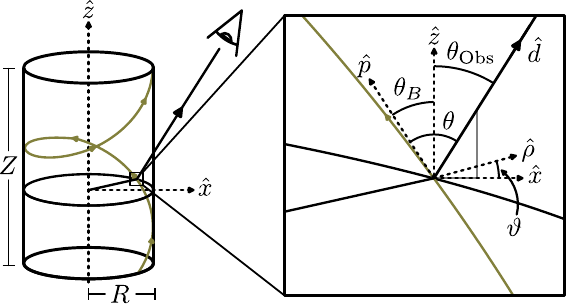}
    \caption{\label{fig: obs flux figure}Definitions of angles used to determine the observed photon flux from a cylindrical zone with a helical magnetic field. The local momentum direction is aligned to the magnetic field, which is at pitch angle of $\theta_B$ to the cylindrical $z$-axis, and the observer is at an angle of $\theta_\text{Obs}$ to the cylindrical $\hat{z}$ axis in the $xz$-plane.}
\end{figure}

For an observer at a distance $d$ and angle $\theta_\text{Obs}$ away from a cylindrical zone of radius $R$ and length $Z$ (see \fref{fig: obs flux figure}), the observed photon flux $F_{p}$, where $p$ is the photon momentum, may be calculated by integrating the flux of photons escaping the $\rho=R$ boundary in the direction of the observer \citep{RybickiLightman_2004}:
\begin{equation}\label{eqn: obs flux int a}
    F_p = \frac{Rc^2}{4\pi d^2}\int_0^Z\mathrm{d}z\int_{-\pi/2}^{\pi/2}\mathrm{d}\vartheta p^3 \left.f(t,R,\vartheta,z,p,u,\phi)\right|_{u,\phi\rightarrow \theta_\text{Obs}}\hat{d}\cdot\hat{\rho},
\end{equation}
where the distribution function $f(\boldsymbol{x},\boldsymbol{p})=f(t,R,\vartheta,z,p,u,\phi)$ has dimensions of momentum$^{-3}\times$length$^{-3}$ and it has been assumed that $d\gg Z\gg R$ such that only the photons passing through the $\rho=R$ surface contributes to the total observed flux, the observer's angle $\theta_\text{Obs}$ is constant over the cylindrical surface, and the full half-surface of the cylinder can be viewed.
For the inner product $\hat{d}\cdot\hat{\rho}=\sin\theta_{Obs}\cos\vartheta$ and a helical magnetic field with fixed pitch angle of $\theta_B$, the local momentum space may be related to the observer's angle via 
\begin{equation}\label{eqn: local theta to obs angle}
    u=\cos\theta=\cos\theta_\text{Obs}\cos\theta_B-\sin\vartheta\sin\theta_\text{Obs}\sin\theta_B.
\end{equation}
Assuming the photon distribution function has no momentum-space $\phi$ angle dependence in the distribution, the integral \eref{eqn: obs flux int a}, can be rephrased using \eref{eqn: local theta to obs angle} as an integral over the local momentum-space angle $u$ rather than $\vartheta$:
\begin{equation}\label{eqn: obs flux int b}
    F_p = \frac{RZc^2}{4\pi d^2\sin\theta_B}\int_{u_-}^{u_+}\mathrm{d}u\, p^3 f(t,p,u),
\end{equation}
with $u_\pm=\cos(\theta_\text{Obs}\mp\theta_B)$ and assuming $f$ has no dependence on $R,\vartheta,z$, which is true for a single zone.

If $\theta_B=0$, the local momentum space angle $\theta$ is equivalent to the global cylindrical angle $\vartheta$ and instead of \eref{eqn: obs flux int b}, the observed spectral energy distribution is given by: 
\begin{equation}
    F_p = \frac{2RZc^2}{4\pi d^2}p^3 f(t,p,\cos\theta_\text{Obs})\sin\theta_\text{Obs}.
\end{equation}
The photon flux in momentum can then be related to the more standard flux in frequency by $F_\nu=F_p\frac{\mathrm{d}p}{\mathrm{d}\nu}=\frac{h}{c}F_p$ and the spectral energy distribution $\nu F_\nu=pF_p$.

